\pdfoutput=1
\documentclass[preprint2]{aastex}

\slugcomment{Received...;Accepted}

\shorttitle{The SIMES survey}
\shortauthors{Baronchelli et al.}

\usepackage{graphicx}
\newcommand{\mic}{$\mu$m}
\DeclareGraphicsExtensions{.pdf,.png,.jpg}

\begin{document}

\title{The \emph{Spitzer}-IRAC/MIPS Extragalactic survey (SIMES) in the South Ecliptic Pole field}

\author{I. Baronchelli\altaffilmark{1, 2}, C. Scarlata\altaffilmark{1}, G. Rodighiero\altaffilmark{2}, A. Franceschini\altaffilmark{2}, P. L. Capak\altaffilmark{3}, S. Mei\altaffilmark{3,4,5},   M. Vaccari\altaffilmark{6,7}, L. Marchetti\altaffilmark{8}, P. Hibon\altaffilmark{9}, C. Sedgwick\altaffilmark{8}, C. Pearson\altaffilmark{8,10,11}, S. Serjeant\altaffilmark{8}, K. Men\'endez-Delmestre\altaffilmark{12}, M. Salvato\altaffilmark{13}, M. Malkan\altaffilmark{14}, H. I. Teplitz\altaffilmark{3}, M. Hayes\altaffilmark{15}, J. Colbert\altaffilmark{3}, C. Papovich\altaffilmark{16}, M. Devlin\altaffilmark{17}, A. Kovacs\altaffilmark{1,3}, K. S. Scott\altaffilmark{18}, J. Surace\altaffilmark{3}, J. D. Kirkpatrick\altaffilmark{3}, H. Atek\altaffilmark{19}, T. Urrutia\altaffilmark{20}, N. Z. Scoville\altaffilmark{3}, T. T. Takeuchi\altaffilmark{21} }


\altaffiltext{1}{MN Institute for Astrophysics, University of Minnesota, 116 Church St. SE,  Minneapolis, MN 55455, USA}
\altaffiltext{2}{Dipartimento di Fisica e Astronomia, Universit${\grave{a}}$ di Padova, vicolo Osservatorio, 3, 35122 Padova, Italy}
\altaffiltext{3}{California Institute of Technology, 1200 E. California Blvd., Pasadena, CA, 91125, USA}
 \altaffiltext{4}{GEPI, Observatoire de Paris, PSL Research University,  CNRS, University of Paris Diderot, 61, Avenue de l'Observatoire 75014, Paris France}
\altaffiltext{5}{University of Paris Denis Diderot, University of Paris Sorbonne Cit\'e (PSC), 75205 Paris Cedex 13, France}
\altaffiltext{6}{Astrophysics Group, Physics Department, University of the Western Cape, Private Bag X17, 7535 Bellville, Cape Town, South Africa}
\altaffiltext{7}{INAF - Istituto di Radioastronomia, via Gobetti 101, 40129 Bologna, Italy}
\altaffiltext{8}{Department of Physical Sciences, The Open University, Milton Keynes, MK7 6AA, UK}
\altaffiltext{9}{Gemini South Observatory, Casilla 603, La Serena, Chile} 
\altaffiltext{10}{RAL Space, Rutherford Appleton Laboratory, Chilton, Didcot, Oxfordshire OX11 0QX, United Kingdom}
\altaffiltext{11}{Oxford Astrophysics, Denys Wilkinson Building, University of Oxford, Keble Rd, Oxford OX1 3RH, UK}
\altaffiltext{12}{Observató\'orio do Valongo, Universidade Federal de Rio de Janeiro, Rio de Janeiro, Brazil}
\altaffiltext{13}{Max Planck institute for extraterrestrial Physics, Giessenbachstr. 1, Garching, D-85748, Germany}
\altaffiltext{14}{Department of Physics and Astronomy,UCLA, Physics and Astronomy Bldg., 3-714, LA CA 90095-1547, USA}
\altaffiltext{15}{Department of Astronomy, Oskar Klein Centre, Stockholm University, AlbaNova University Centre, SE-106 91 Stockholm, Sweden}%
\altaffiltext{16}{Department of Physics and Astronomy, Texas A\&M University, College Station, TX 77843-4242, USA }
\altaffiltext{17}{Department of Physics and Astronomy, University of Pennsylvania, 209 South 33rd Street, Philadelphia, Pennsylvania 19104, USA}
\altaffiltext{18}{National  Radio  Astronomy  Observatory,  520  Edgemont  Rd,  Charlottesville, VA, 22903, USA}
\altaffiltext{19}{Laboratoire d'Astrophysique, Ecole Polytechnique F\'ed\'erale de Lausanne, Observatoire de Sauverny, CH-1290 Versoix, Switzerland }
\altaffiltext{20}{Leibniz Institut f\"ur Astrophysik Potsdam, An der Sternwarte 16, D-14482 Potsdam, Germany}
\altaffiltext{21}{Division of Particle and Astrophysical Science, Nagoya University, Furo-cho, Chikusa-ku, Nagoya 464-8602, Japan} 

\begin{abstract} We present the \emph{Spitzer}-IRAC/MIPS Extragalactic survey (SIMES) in the South Ecliptic Pole (SEP) field.  The large area covered (7.7 deg$^2$), together with 
 one of the lowest Galactic cirrus emissions in the entire sky and a very extensive coverage by \emph{Spitzer, Herschel, Akari,} and \emph{GALEX}, make the SIMES field ideal for extragalactic studies. The elongated geometry of the SIMES area ($\approx$4:1), allowing for a significant cosmic variance reduction, further improves the quality of statistical studies in this field. Here we present the reduction and photometric measurements of the \emph{Spitzer}/IRAC data. The survey reaches a depth of 1.93 and 1.75 $\mu$Jy (1$\sigma$) at 3.6 and 4.5 \mic, respectively. We discuss the multiwavelength IRAC--based catalog, completed with optical, mid-- and far--IR observations. We detect 341,000 sources with F$_{3.6\mu m} \geq 3\sigma$. Of these, 10\% have an associated 24 \mic\ counterpart, while 2.7\% have an associated SPIRE source. We release the catalog through the NASA/IPAC Infrared Science Archive (IRSA).
Two scientific applications of these IRAC data are presented in this paper: first we compute integral number counts at 3.6 \mic. 
Second, we use the [3.6]--[4.5] color index to  identify galaxy clusters at z$>$1.3. We select 27 clusters in the full area, a result consistent with previous studies at similar depth.
\end{abstract}


\section{INTRODUCTION}
\label{introduction} 

Contrary to the expectation that galaxy formation would proceed via merger-driven bursts of star formation (SF), evidence is now overwhelmingly showing that the bulk of  SF in the Universe happened in a "quiescent" mode, at average rates increasingly higher  at earlier cosmic times \citep[e.g., ][]{2007ApJ...660L..47N,2007ApJ...670..156D,peng2010,2011ApJ...739L..40R}. Although short-lived powerful merger-driven starbursts (SFR$>1000 M_{\odot}$ yr$^{-1}$) do not contribute significantly to the slow process of galaxy growth, they may, however, represent a critical phase in the structural transformation and quenching of the most massive galaxies \citep[e.g., ][]{2011ApJ...739L..40R, 2013ApJ...778..126B}. 

The most actively SF galaxies  at any redshifts tend to be also the most dust-obscured objects. They disappear at rest-frame UV wavelengths and emit most of their energy in the far--IR, where they can be easily identified through imaging between 24 and 500 \mic.
The \textit{Herschel} satellite with its PACS and SPIRE instruments \citep{2010A&A...518L...1P, 2010A&A...518L...2P,2010A&A...518L...3G} have revolutionized the field, producing large samples of mid--IR bright galaxies, selected up to very large redshifts via their bolometric luminosity 
\citep[e.g., ][]{2010A&A...518L..25R, 2011ApJ...739L..40R, 2011ApJ...742...96W, 2012MNRAS.424.1614O, 2012A&A...539A.155M, 2012A&A...545A..45R, 2013MNRAS.432...23G, 2014A&A...562A..30S}.
 To understand the physical nature of these sources, however, detailed sampling of the full spectral energy distribution and (spectroscopic/photometric) redshifts are needed, and require a secure counterpart association at shorter wavelengths. With secure counterparts, and sufficient ancillary data, accurate photometric redshifts can be computed which will allow the effective use of the Atacama Large Millimeter/submillimeter Array (ALMA), not only to measure spectroscopic redshifts, but also to understand the physical conditions of the molecular gas reservoir in these objects.  

In this paper we present the   \textit{Spitzer}---IRAC/MIPS Extragalactic survey (SIMES),  an infrared survey carried out with the \textit{Spitzer} Space telescope \citep{2004ApJS..154....1W,2004ApJS..154...10F} of a 7.7 deg$^{2}$ field close to the South Ecliptic Pole (SEP) at 3.6 and 4.5 $\mu$m.  The SIMES field, centered at ($\alpha , \delta$) = ($4^{\mathrm{h}}44^{\mathrm{m}}$, -53$^{\circ}$30'), has among the lowest Galactic cirrus emission in the entire sky  \citep[$\sim$2-3 MJy str$^{-1}$ at 100 $\mu$m,][]{1998ApJ...500..525S,2006PASJ...58..673M}, thus minimizing the extinction in the UV and optical bands as well as maximizing the sensitivity at far-IR wavelengths. 
This field is therefore very appealing for full multiwavelength exploitation.
Furthermore, it has the unique advantage of having an elongated geometry (axial ratio of approximately $\sim 4:1$), which minimizes the cosmic variance compared to square fields of similar depth and area on the sky \citep{2008ApJ...676..767T}.

The SIMES field has been the target of a vast array of multiwavelength observing programs from major observatories: \emph{Spitzer} \citep{2011MNRAS.411..373C}, \emph{Herschel} \citep{2012MNRAS.424.1614O, 2014MNRAS.444.2870W}, \emph{GALEX} \citep{2007ApJ...655..863D}).
In particular, together with the North Ecliptic Pole (NEP) area, it is one of the two fields including the deepest contiguous observations by the \emph{Akari} IR observatory  in the context of the \emph{AKARI} Deep Field South survey \citep[ADFS,][]{2006PASJ...58..673M,2011ApJ...737....2M}. These observations provide us with the most extensive photometric coverage in the mid--IR available for cosmological surveys, of particular relevance for the analysis of dust-obscured active galaxies and AGNs. 
Very important for the identification at IR and optical wavelengths is the availability of \textit{Spitzer} 24 \mic\ imaging \citep{2004ApJS..154...25R}, which --with its 5\farcs0 beam-- nicely links imaging at shorter and  longer wavelengths.
Until our survey, however,
the SIMES field was missing the crucial IRAC coverage required to associate the majority of broad-beamed/confused $\geq 24$ \mic\ sources with physically understood astrophysical objects (detected at $\lambda <1$ \mic).  Here we present the new IRAC observations of the SIMES field,  targeting this outstanding   wavelength gap and thus allowing the full exploitation of the available longer wavelength data. These data will also be crucial for the measurement of physical properties of the high redhisft objects, including their photometric redshifts and stellar masses.

The paper is organized as follows. We describe the IRAC observations and catalog preparation in Section \ref{IRAC_cat}.  The matching with the long wavelength ancillary data is described in Section \ref{MW_DATA_CATALOG_SECTION}. Finally we discuss initial results on the 3.6 \mic\ number counts and the identification of intermediate redshift galaxy clusters, in Section \ref{sec:results}. Throughout the paper we assume a standard flat cosmology with H$_{0}$=70 Km s$^{-1}$ Mpc$^{-1}$, $\Omega_{M}=0.3$ and $\Omega_{\Lambda}=0.7$. Wherever magnitudes or colors are reported, the AB magnitude system is implicitly assumed.

\section{OBSERVATIONS AND DATA ANALYSIS}
\label{IRAC_cat}
\begin{figure*}[ht!]
\centering
\includegraphics[width=15.0cm]{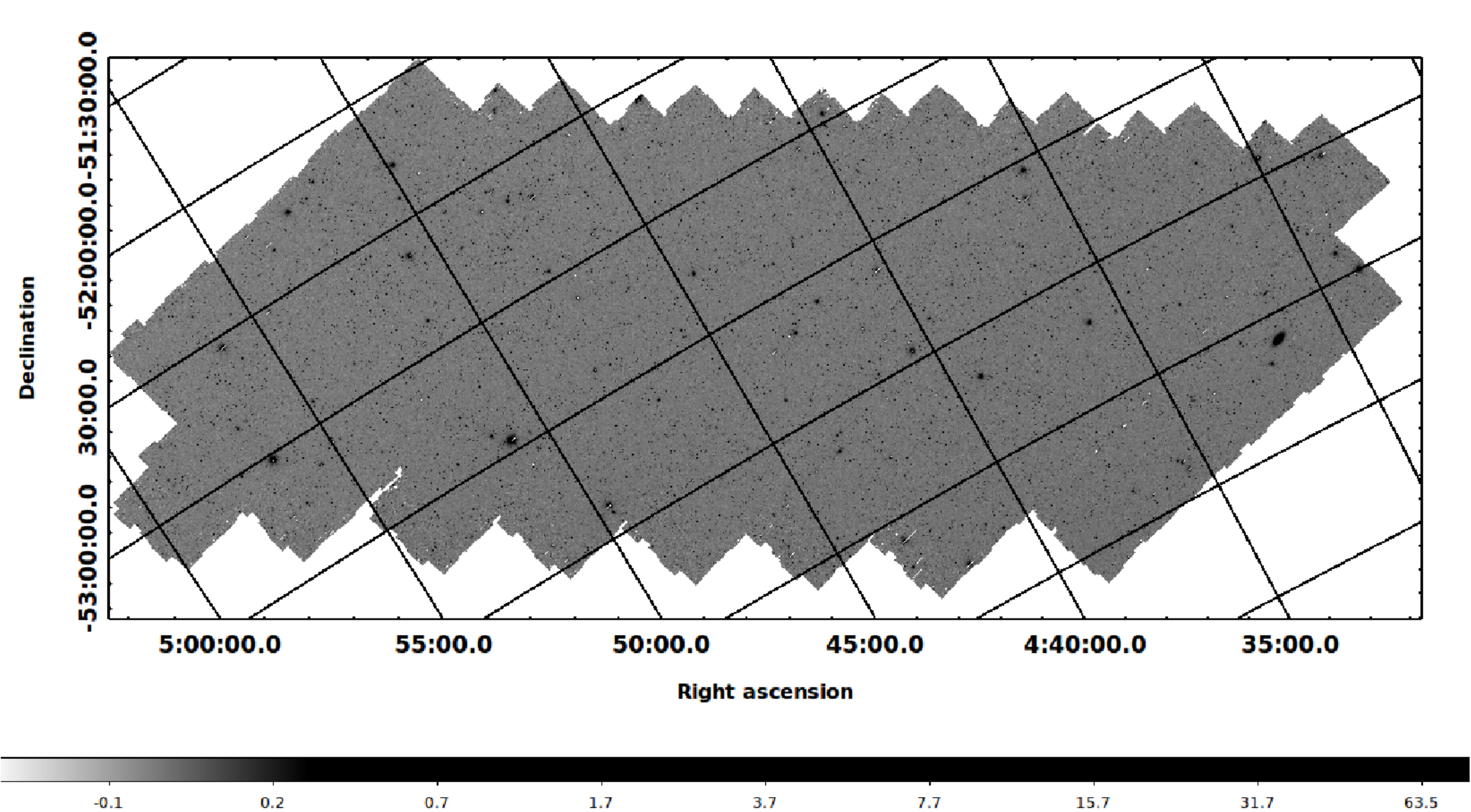}
\includegraphics[width=15.0cm]{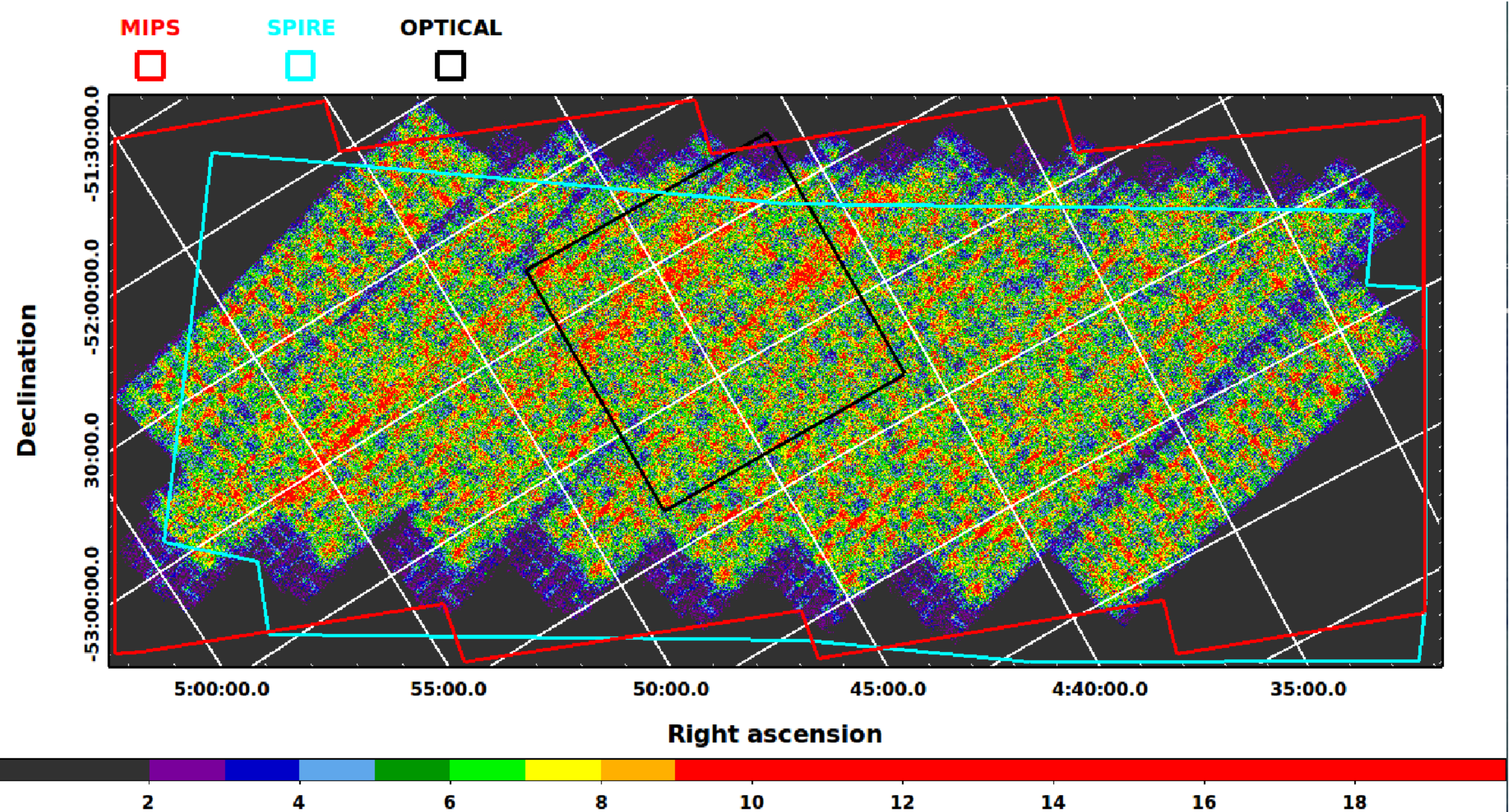}
\caption{{\bf Top panel} IRAC 3.6 \mic\ mosaic of the SIMES field.  The color scale is  in units of MJy sr$^{-1}$. {\bf Bottom panel} Coverage map at 3.6 \mic. The color scale shows the number of frames per pixel.  The areas covered by MIPS, SPIRE, and WFI--R$_{c}$ (optical) are shown in red, cyan, and black respectively.}
\label{IRAC1_img}
\end{figure*}

SIMES is a \emph{Spitzer} Cycle~8  
General Observer program (PID 80039, P.I.: Scarlata) observed during the warm mission phase. The SIMES survey in the IRAC bands was designed with the goal of complementing the existing MIPS 24 \mic\ and far--IR observations. The  7.74 deg$^{2}$  field was covered in 78 hours with the IRAC instrument in both channel 1 and channel 2, corresponding to imaging at 3.6 \mic\ and 4.5 \mic, respectively. In order to efficiently cover the elongated  region, we used  multiple $4\times 16$ IRAC AORs. This strategy was chosen to minimize the effect of the substantial rotation (1 deg per day) at the field latitude.
The field was covered in two visits, between 2011 November 16$^{\mathrm{th}}$ and 23$^{\mathrm{rd}}$, in order to facilitate identification and removal of asteroids. The first and the second visits consisted of 3$\times$30 s and 2$\times$30 s frames respectively, obtained with a medium cycling dither pattern, for a total exposure time of 150 s.
The reduction of the IRAC data generally followed the procedure used by the Spitzer Enhanced Imaging Products (SEIP) mosaic pipeline \citep{2013AAS...22134006C} with one additional step.  Before the SEIP mosaic processing a median image was created for each AOR (observing block) and subtracted from the frames to remove residual bias in the frames and persistence from previous observations.  The MOPEX Overlap routine \citep{2005PASP..117..274M} was then used on the background subtracted images to remove any residual background variation from frame to frame.  The median subtracted frames were then combined with the MOPEX mosaic pipeline \citep{2005ASPC..347...81M}.  The outlier, and box-outlier modules were used to reject cosmic rays, transient, and moving objects. 
Many of the exposures were affected by latent images from prior observations of a bright object, although the impact was much more severe in channel 1 than channel 2, where it effectively doubled the noise over the background limited estimate.  The effects of these latent images could not be fully mitigated because they faded during the observations, so a perfect model could not be produced. As a result, the 3.6 \mic\ data have a sensitivity comparable to the 4.5 \mic\  (see Section~\ref{SNR_section} and Table~\ref{coverage_table}).
As a final step, the data were then interpolated onto a 0\farcs6 pixel scale using a linear interpolation and combined with an exposure time weighted mean combination. A mean, median, coverage, uncertainty, and standard-deviation image were created.  
The final resulting mosaic is shown in the top panel of Figure~\ref{IRAC1_img}, while  the bottom panel shows a comparison of  the IRAC coverage with the coverage in the MIPS, SPIRE, and optical surveys within the same area.

\subsection{Source Extraction and Photometry}
\label{Sextr_phot}
For the detection and extraction of sources we used
the \emph{SExtractor} software \citep{1996A&AS..117..393B} in dual
image mode, using the 3.6 \mic\ map as the detection image, and the uncertainty map as a weight image. During the detection step, we used a local background calculated over an area of 32$\times$32 pixels filtered with a 3--pixel size top--hat kernel. 
We set a 1.5$\sigma$ threshold, with a minimum of 5 connected pixels above the background noise. For each object we computed AUTO total fluxes, as well as aperture fluxes measured in apertures of 4\farcs8, 7\farcs2, and 12\farcs0 diameter. 

\emph{SExtractor} AUTO fluxes are  estimates of the total flux of a source in an elliptical aperture with semi-major axis ($a$) proportional to the Kron radius of the object \citep[$R_K$,][]{1980ApJS...43..305K}. We chose $a=2.5\ R_K$, by setting the \emph{SExtractor} parameter Kron\_fact $=2.5$\footnote{In \emph{SExtractor} nomenclature, $a$$=$KRON\_RADIUS $\times$ A\_IMAGE, where A\_IMAGE is the luminosity profile RMS, in pixels, along the major axis direction, while KRON\_RADIUS=Kron\_fact$\times R_K$, with $R_K$ in units of A\_IMAGE.}. This choice ensures that the aperture includes more than the 90\% of the total galaxy flux\footnote{The Kron aperture includes a different fraction of the total light of a galaxy, depending on the value of the S\'ersic index \emph{n} of its surface brightness profile \citep{1963BAAA....6...99S,2005PASA...22..118G}} \citep{1980ApJS...43..305K}. For apertures with $a< 3.5$ pixels (2\farcs1), the AUTO flux is computed within a circular aperture.
AUTO fluxes account for the real apparent dimension of each source, the elliptical shape of the observed  isophotes and the source's radial surface brightness profile. 

\begin{figure}
\centering
\includegraphics[width=8.0cm]{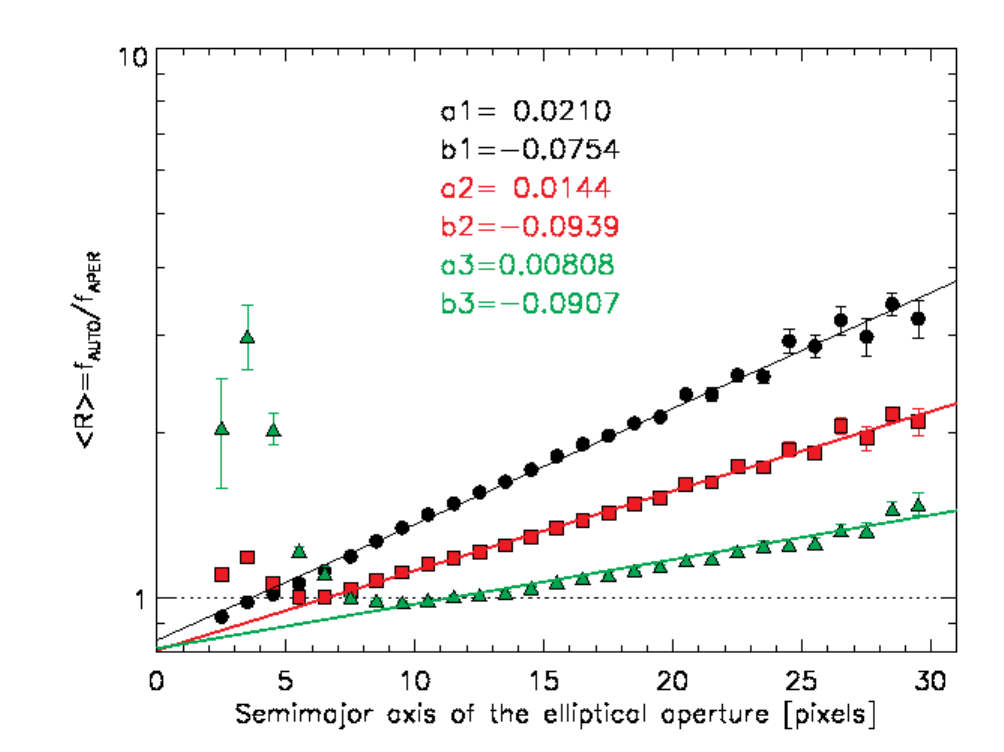}
\caption{ \label{aperture_corrections} Average ratio between AUTO fluxes and
      uncorrected aperture fluxes (4\farcs8 in black, 7\farcs2 in red,
      12\farcs0 in green), as a function of the semi-major axis of the
      Kron elliptical aperture, together with best linear fit coefficients, expressed as $\log \langle R \rangle=a\times$(semi-major axis)$+b$. The figure shows how an increasing amount of the source's emission is missed when using fixed circular apertures to compute fluxes.}
\end{figure}

In Figure~\ref{aperture_corrections} we report the mean ratio $\langle R \rangle$ between the AUTO fluxes and the APERTURE fluxes computed for our sources in bins of semi-major axis.  $\langle R \rangle$  increases with the increasing apparent dimension of the sources indicating that a fraction of the source€™s' emission is missed when using a fixed aperture. There is a strong linear correlation between $\log\langle R \rangle$ and the semi-major axis of the elliptical aperture, with different slopes depending on the size of the circular aperture used. 
The ratio becomes $\sim$1 when the semi major axis of the elliptical aperture has a dimension similar to that of the circular aperture used for the comparison. When considering the smallest elliptical apertures and the largest circular aperture, we observe a large deviation from $\langle R \rangle\sim 1$. This effect is likely due to an overestimate of the background that becomes appreciable when the elliptical and circular apertures have very different sizes. Sources characterized by small apparent dimensions (i.e. small semi-major axis of the elliptical aperture) tend to have smaller aperture fluxes for larger aperture sizes (green curves in Figure~\ref{aperture_corrections}).
Thus, hereafter, all fluxes reported are total fluxes ``FLUX\_AUTO'' measured within the Kron \emph{SExtractor} apertures.

\subsection{Survey Sensitivity}
\label{SNR_section}

\begin{figure}[!ht]
\begin{center}
\includegraphics[width=7.5cm]{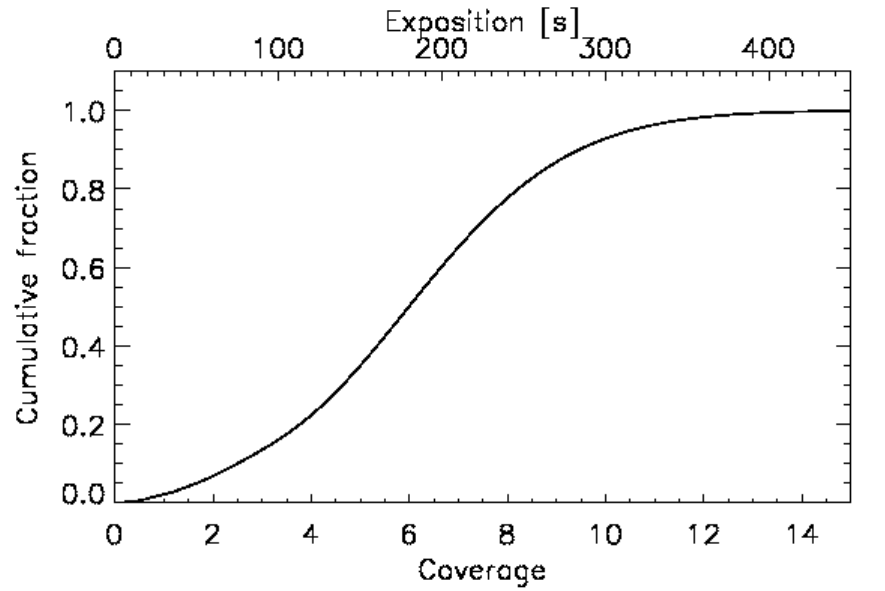}
\caption{\label{cov_distrib}Cumulative distribution of the
  pixel coverage. More than 70\% of the pixels in the mosaic are at the nominal 150 s coverage.}
\end{center}
\end{figure}

The mapping strategy adopted to cover the large SIMES area results in a  varying coverage across the field, with a resulting noise variation. Figure~\ref{cov_distrib} shows the cumulative distribution of the pixel coverage in the 3.6 \mic\ mosaic, where the coverage on the horizontal axis is defined as the number of 30 s exposures per pixel. More than 70\% of the mosaic is at or above the planned coverage of 150 s.

In order to identify reliable detections in the 3.6 \mic\ catalog, we follow \citet{2005AAS...207.6301S} and compute a coverage-based signal--to--noise ratio for each source. 
\citet{2005AAS...207.6301S} compute the noise 
$\sigma$ corresponding to the mean coverage within the mosaic  ($\langle C\rangle$, where C is the number of exposures per pixel) and then scale it by a factor $f$ that accounts for the specific coverage, $C$, of the aperture used for the flux measurement, i.e., $f=\sqrt{\langle C\rangle/C}$ . This procedure assumes that the noise scales as the square root of the exposure time. 
However, the noise contribution from faint unresolved sources could be substantial in the deepest regions of the mosaic.

In order to check the $t_{exp}^{-0.5}$ assumption, we empirically derived the noise properties of the mosaic as a function of the actual coverage. 
First, we measured the flux in 8 pixel diameter apertures distributed in an homogeneous grid covering the mosaic.
Then, we divided our measures in different groups, according to the coverage underlying the apertures in which they were obtained.
In each  bin of coverage, we fitted a Gaussian function to the distribution of aperture fluxes, symmetrized with respect to the median to include only background dominated apertures. The standard deviation of the best-fit Gaussian distribution in each bin of coverage is then a measurement of the average background noise corresponding to that coverage.  

To check the normal distribution of the pixels noise, we applied both the Kolmogorov--Smirnov (K--S) and the Anderson--Darling (A--D) tests to the negative side of the pixel flux distribution. Because the noise level is expected to depend on the exposure time, we consider seven equally sized bins of coverage ranging between two and ten. At the nominal coverage of five, using the K--S test, the probability to find the computed difference \emph{D}=3.9$ \times 10^{-3}$ between the cumulative distribution of fluxes and that expected from a normal distribution is 0.74. Using the A--D test, we found a difference A$^{2}$=0.55, that is close to the reference value of 0.576, corresponding to the rejection of the null hypothesis (normality) with a 15\% level of significance. At lower and higher coverages, the probability from the K--S test ranges from 0.28 to 0.96, while A$^{2}$ ranges from 0.49 to 1.87, indicating that when a deviation from the normal distribution is present, it is small.

\begin{figure}[!ht]
 \begin{center}
\includegraphics[width=7.5cm]{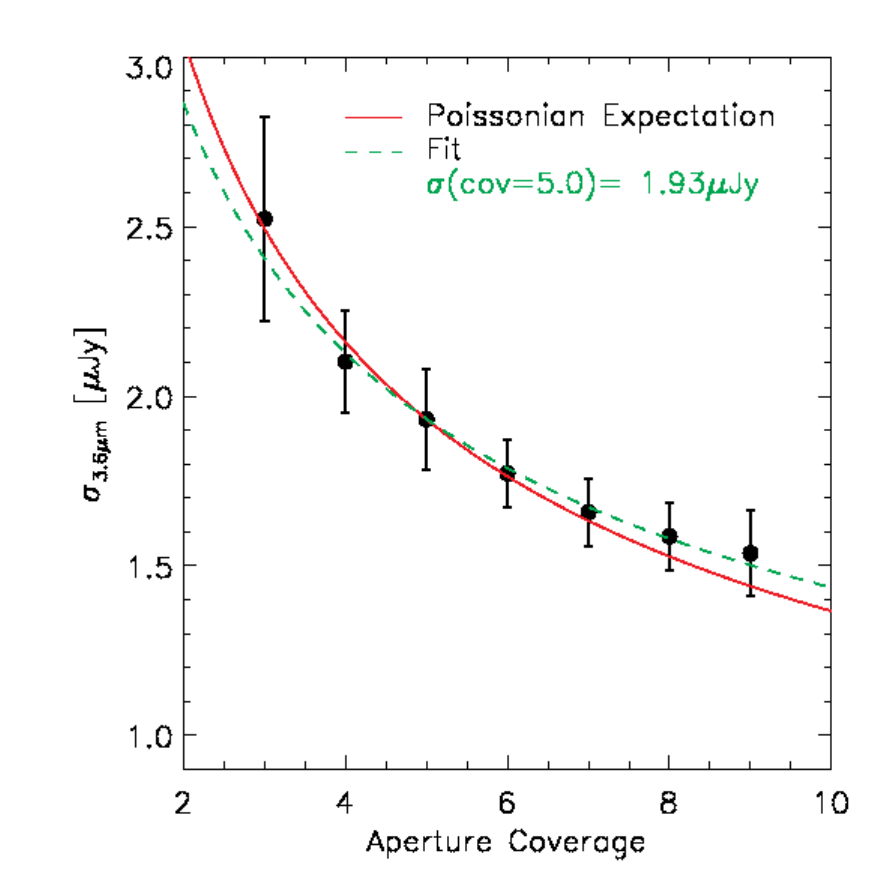}
\caption{\label{SNR_cov} Sky background noise, $\sigma$, as a function of aperture coverage (see text for details).  The expected trend for background-dominated noise is shown with a red solid line, while the observed best fit relation is shown with a green dashed line. We cut the IRAC 3.6 $\mu$m catalog at a 3$\sigma$ level, where $\sigma$ is estimated from the average coverage of each source. }
\end{center} 
\end{figure}

In Figure \ref{SNR_cov}, we show the resulting $\sigma$ as a function of average coverage $<$C$>$ in each of the six bins. The dependency with $C$ is in agreement with the Poissonian approximation (shown with a continuous red line). The theoretical curve is normalized at the nominal coverage of the survey (C=5). The trend between $\sigma$ and $C$ is well reproduced by the relation $\sigma \propto C^{-\alpha}$, with $\alpha=0.43\pm 0.09$ (green dashed line in Figure~\ref{SNR_cov}).
To calculate the signal--to--noise ratios, we computed APERTURE flux and noise in the same aperture (4\farcs8 diameter). For each source, the underlying coverage is computed as the median value in the aperture. We retain in the final catalog only sources with 3.6 \mic\ flux above 3$\sigma$. The final IRAC-based catalog constructed in this way includes 341006 sources.

We compute the sensitivity of the 4.5 \mic\ observations using the same method described for the 3.6 \mic\ data. Again, measuring $\sigma$ as a function of the average coverage we found an agreement with the Poissonian expectation, with $\alpha(4.5\ \mu m)=0.53\pm 0.08$.

At all coverages, we found a 4.5 \mic\ depth $\sigma$ comparable or smaller than that measured at 3.6 \mic. In particular, at the nominal coverage of the survey we measure $\sigma$=1.93 $\mu Jy$ at 3.6 \mic\ and $\sigma(C=5)$=1.75 $\mu Jy$ at 4.5 \mic. As noted before, this  behavior is due to the effect of latent images being more pronounced  in the 3.6 \mic\ than in the 4.5 \mic\  channel.

\subsection{Survey Completeness and Contamination}
\label{sec:completeness}
We estimated the survey completeness as a function of the 3.6 \mic\ flux, adding artificial sources to the original IRAC mosaic and extracting them with the same procedure used for the real IRAC map. To create the artificial sources, we generated a synthetic PSF using the median of 76 images of point sources extracted from the original 3.6 \mic\ map.
These sources were selected for being isolated (closest counterpart distance $>$22\farcs0) and with fluxes near 100 $\mu$Jy. 
Moreover, we excluded clearly extended sources, characterized by KRON\_RADIUS$\times$A\_IMAGE$>$10 pixels and sources located at the edge of the map.

\begin{figure}[!ht]
 \begin{center}
\includegraphics[width=7.5cm]{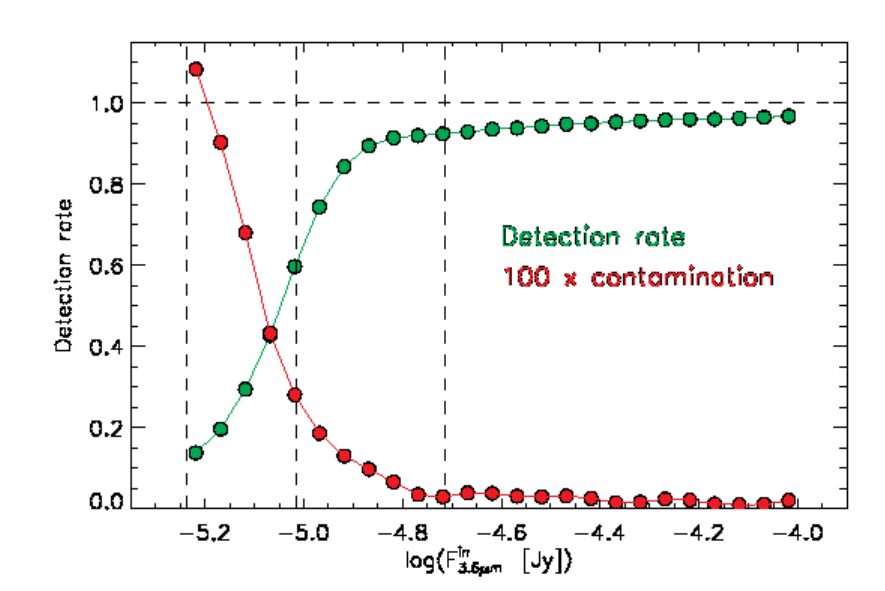}
\includegraphics[width=7.5cm]{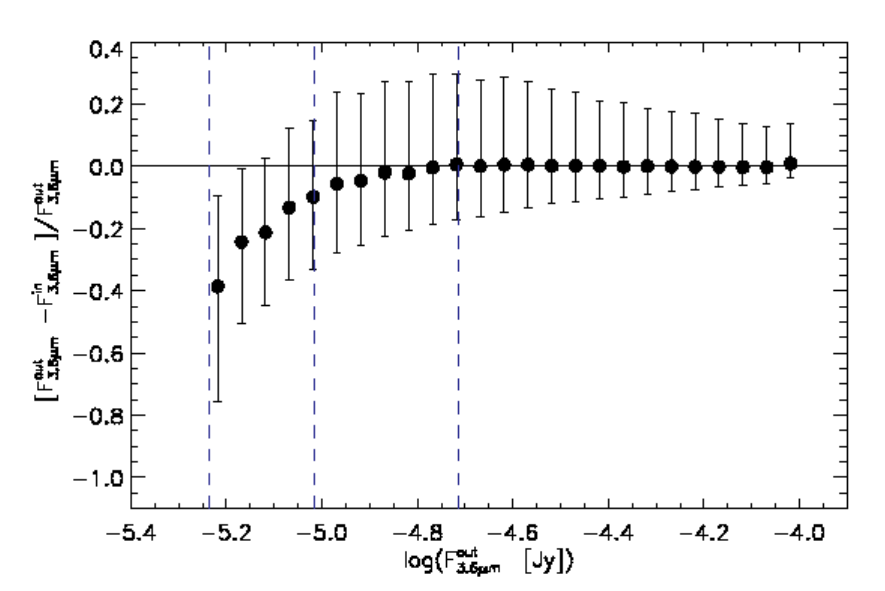}

\caption{\label{fig:completeness} {\bf Top panel} Completeness and contamination computed through numerical simulations.  The completeness level drops to approximately 50\% at 3.6 \mic\ flux $\sim$ 9 $\mu$Jy. The contamination rate (multiplied by a factor of 100, for clarity) reaches a maximum of 1.1\% at the 3$\sigma$ flux. {\bf Bottom panel} Flux accuracy of the simulated sources.
The error bars represent the 1$\sigma$ range of flux accuracy resulting from the simulations.
 Source fluxes start to become systematically underestimated (by more than 10\%) below $\sim$ 9 $\mu$Jy. The vertical dashed lines show our 3, 5, and 10 $\sigma$ flux limits.}

\end{center} 
\end{figure}

We simulated approximately 69,000 artificial sources with 31 different 3.6 \mic\  fluxes in the range $\sim$3--100 $\mu$Jy. For each of the 31 groups, we simulated an independent IRAC map, randomly distributing 2233 same--flux artificial sources along with the real ones. After the extraction, we computed the detection rate (i.e., the completeness) as the ratio between the number of sources inserted in the map and the recovered ones. This approach allows us to maximize the number of sources inserted in the maps without artificially increasing the spatial density of the sources, as would happen if we added all the simulated sources at once.
 The results of this analysis are presented in Figure~\ref{fig:completeness}, where we show that the completeness drops below 50\% at 3.6 \mic\ flux of approximately 9.0 $\mu$Jy (corresponding to a source flux of approximately $4.7\sigma$).

We also investigate the flux accuracy as a function of the artificial source flux, by computing the average difference between the flux of the simulated sources and their flux after the extraction. The results, presented in the bottom panel of Figure~\ref{fig:completeness}, show that the accuracy of the flux measurements is a function of the output fluxes. 
For sources at a 3$\sigma$ level, the recovered flux ranges (1$\sigma$ of the data distribution) from $\sim$75\% to $\sim$10\% below the input flux, while for sources at the 10$\sigma$ level, the range is from $\sim$20\% below to $\sim$30\% above the input flux.
There is an indication that faint sources ($F_{3.6\mu m }<5\sigma$) have systematically underestimated fluxes, although the scatter in this range is also larger. 
The detection rate as a function of 3.6 \mic\ flux is summarized in Table~\ref{completenessI1_table}. These values have been used in computing the 3.6 \mic\ number counts presented in Section~\ref{Counts}.

We estimate the false positive rate (i.e. contamination rate) applying the same extraction technique and 3$\sigma$ cut used for the original 3.6 \mic\ map, to the inverted 3.6 \mic\ image (pixel fluxes multiplied by -1). We obtain a total contamination rate of 0.21\%. In Figure~\ref{fig:completeness}, the contamination (multiplied for a factor of 100) is shown as a function of the flux of the spurious extracted sources. Spurious sources represent  $\sim$1.1\% of all sources at the 3$\sigma$ level and  $\sim$0.25\% at 5$\sigma$. In order to obtain an independent upper limit to the total contamination rate, we apply the Benjamini--Hochberg test \citep{bh1995} under the assumption that the background distribution is known and Gaussian.  For each flux in the catalog we can then compute its p--value under the null hypothesis that it was extracted from the background population. The p--value relative to the 3$\sigma$ flux threshold adopted for the 3.6\mic\ catalog  (p$_{3\sigma}$) corresponds to a value C(p$_{3\sigma}$) in the p-values cumulative distribution. The expected contamination rate $\epsilon=(1-\alpha)Np_{3\sigma}/$C(p$_{3\sigma}$)$<Np_{3\sigma}/$C(p$_{3\sigma}$), where N is the number of apertures used in the test and $\alpha$ is the ratio between the number of spurious and real sources in the catalog. At the nominal coverage of 5, the upper limit on the contamination rate is then 1.2\%.

\begin{deluxetable}{cc}
\tabletypesize{\footnotesize}
\tablecolumns{2}
\tablewidth{0pc}
\tablecaption{Completeness as a Function of 3.6 $\mu\mathrm{m}$ Flux}
\tablehead{
\colhead{Completeness}    & \colhead{Flux 3.6 \mic} \\
\colhead{(\%) } & \colhead{($\mu$Jy)}      }
\startdata
50  & 9.0 \\ 
75  & 10.8 \\
90  & 14.1 \\
95  & 40.0 \\
97  & 107.1 \\
99  & 224.2 \\
\enddata
\label{completenessI1_table}
\end{deluxetable}


\section{ANCILLARY DATA}
\label{MW_DATA_CATALOG_SECTION}

The SIMES field is fully covered  with both \emph{Spitzer} MIPS (24 and 70 $\mu$m), and \emph{Herschel} SPIRE (250, 350, and 500 $\mu$m) as well as by \emph{Akari}. In the present paper we report on the MIPS and SPIRE observations, while the \emph{Akari} data will be discussed in a forthcoming paper (I. Baronchelli et al. in preparation). The central square degree is also covered by optical imaging (see Figure~\ref{IRAC1_img}).
In the following sections we describe how we merged the IRAC-based catalog with the publicly available MIPS \citep{2011MNRAS.411..373C} and SPIRE \citep[\emph{HerMES}, DR2,][]{2010MNRAS.409...48R,2012MNRAS.424.1614O, 2012MNRAS.419..377S,2014MNRAS.444.2870W} catalogs. Section~\ref{WFI_R_SECT} describes the data reduction, photometry, and matching of the optical data. The main properties of the multi--wavelength data are summarized in Table~\ref{coverage_table}.

\begin{deluxetable}{llcccccc}
\tabletypesize{\footnotesize}
\tablecolumns{7}
\tablewidth{0pc}
\tablecaption{Available Ancillary Data}
\tablehead{
\colhead{Band} &\colhead{Instrument}   & \colhead{Overlap Area \tablenotemark{a}} & \colhead{Depth} & \multicolumn{4}{c}{Number of Identified Counterparts \tablenotemark{b}}  \\
\colhead{(\mic)}  & & \colhead{(deg$^{2}$)}  & \colhead{} & \colhead{All} & \colhead{MIPS 24} & \colhead{SPIRE} & \colhead{MIPS 24 \& SPIRE} }
\startdata
3.6  & IRAC   & 7.74 & 5.80 $\mu$Jy (3$\sigma$) \tablenotemark{c,d} & 341006 & 25132 & 9447 & 7041 \\ 
4.5  & IRAC  & 7.26 & 5.25 $\mu$Jy (3$\sigma$) \tablenotemark{d}    & 320460 & 23688 & 9320 & 6947 \\
24   & MIPS  & 7.66 & 0.26 mJy (50\% compl.) \tablenotemark{e}      & 25132 (60) & 25132 (60)& 7041 (60)& 7041 (60) \\
70   & MIPS  & 7.66 & 24 mJy   (50\% compl.) \tablenotemark{e}      & 882  & 882 & 692 & 692 \\
250  & SPIRE & 6.52 & 15.6 mJy (3$\sigma$) \tablenotemark{f}        & 8743 (50) & 6666 (50) & 8743 (50) & 6666 (50) \\
350  & SPIRE & 6.52 & 12.7 mJy (3$\sigma$) \tablenotemark{f}        & 9416 (60) & 7015 (60) & 9416 (60) & 7015 (60) \\
500  & SPIRE & 6.52 & 18.5 mJy (3$\sigma$) \tablenotemark{f}        & 8624 (58) & 6354 (58) & 8624 (58) & 6354 (58)\\
0.65 & WFI   & 1.13 &  0.53 $\mu$Jy (3$\sigma$)                      & 27585 & 2279 & 808 & 680 \\

\enddata
\tablenotetext{a}{Area covered in both the IRAC 3.6 \mic\ band and in the band indicated in the first column.}
\tablenotetext{b}{The additional number of MIPS--SPIRE sources without 3.6 \mic\ counterparts is indicated in parenthesis.}
\tablenotetext{c}{The IRAC 3.6 $\mu$m catalog is cut at a 3$\sigma$ level, as described in the text, keeping into account the underlying coverage for each source. Sources with fluxes below the value reported in the Table can consequently be found in the catalog.} 
\tablenotetext{d}{The value of $\sigma$ reported is estimated for the nominal coverage of the survey (C=5).} 
\tablenotetext{e}{From \citet{2011MNRAS.411..373C}. Minimum 24 $\mu$m flux in the catalog: 0.20 mJy. Minimum for MIPS identified sources with a SPIRE counterpart: 0.31 mJy. } 
\tablenotetext{f}{1$\sigma$ values from \citet{2012MNRAS.424.1614O}. We included in our catalog only the sources with a flux higher then 3$\sigma$ in at least one of the SPIRE bands. } 
\label{coverage_table}
\end{deluxetable}

\subsection{MIPS 24 and 70  $\mu$m}
\label{sec:MIPS}
The MIPS~24 \mic\ catalog is described in \cite{2011MNRAS.411..373C}.
The \citet{2011MNRAS.411..373C} catalog covers an area of $\sim12$ deg$^{2}$ in the south ecliptic pole region and includes the counterparts at 70 $\mu$m of the 24 $\mu$m detected sources, 
so we limit the analysis to
the cross--correlation between IRAC and MIPS--24, and report the
70 $\mu$m association identified in the original MIPS catalog. 
\citet{2011MNRAS.411..373C} estimate that the 24 $\mu$m catalog is 50\%
complete at 0.26 mJy and 80\% complete at 0.32 mJy, while the source
reliability is 96\% at 0.285 mJy.

In order to identify the most likely IRAC counterpart to each
MIPS source, we proceed as follows. For each MIPS-24 $\mu$m source, we
searched the IRAC catalog for the nearest object inside a radius equal
to the quadratic sum of the $\sigma$ of the PSF of the two instruments
(i.e., a search radius of 2\farcs6). In the matching process, we
identified a small systematic shift\footnote{We verified
  that  the shift did not depend on the position in the large mosaic, thus
  indicating that any distortion in the IRAC mosaic was properly
  accounted for.} (of the order of $\Delta RA=$0\farcs099, $\Delta
DEC=$0\farcs49) between the two catalogs. Therefore, we corrected the MIPS
positions before searching for the nearest IRAC counterpart. We report
in the final catalog both the corrected and the original coordinates
of the sources in each band. In Table \ref{counterp_search_table} we
report the distance and the average RA and DEC shifts of all sources
matched in the catalog. When multiple IRAC sources were found within
the search area (see Figure~\ref{Ncnterpts}), we associated the closest IRAC object. 
This happens for 514 MIPS sources that we flag as uncertain identifications (``N\_IRAC\_MIPS''
parameter greater than one). All the other potential IRAC counterparts
can be found in the catalog. Out of all MIPS sources (25132 objects), 98.0\% 
have an unique IRAC counterpart within a region of 2\farcs6 radius. MIPS sources without an IRAC counterpart are generally not included in our catalog. The only exception is represented by 60 visually checked sources in the IRAC covered area with a reliable SPIRE counterpart.

\subsection{SPIRE 250, 350, 500 $\mu$m}
\label{subsec:SPIRE}
The SIMES field was observed as part of the \emph{Herschel} Multi-tiered Extragalactic Survey \citep[HerMES, ][]{2012MNRAS.424.1614O, 2014MNRAS.444.2870W}. The second data release of the SPIRE XID catalogs \citep[DR2, ][]{2010MNRAS.409...48R, 2012MNRAS.419..377S,2014MNRAS.444.2870W} covers approximately 84\% of the field, and includes all sources identified at 1$\sigma$ level at 250, 350 or 500 $\mu$m. We keep here only those sources with fluxes above 3$\sigma$ in at least one SPIRE band. 

The large size of the SPIRE PSF (18\farcs0 at 250 $\mu$m) prevents us from directly cross-correlating the SPIRE and IRAC catalogs. Instead, we exploited the MIPS~24 $\mu$m detections as a ``bridge'' between the two wavelengths, i.e., we searched the SPIRE counterparts given the MIPS--24 prior position. We performed a direct SPIRE-IRAC correlation only when a SPIRE source did not have a MIPS counterpart.  As we did for the MIPS--IRAC correlation, before correlating SPIRE and MIPS counterparts, we corrected the SPIRE coordinates for the small average offset between MIPS and SPIRE positions using the original MIPS coordinates as reference. In our catalog we report both the original SPIRE (RA, DEC) coordinates and the coordinates corrected to the average SPIRE--MIPS and then MIPS--IRAC shifts.

Given the MIPS positions, we searched the SPIRE counterparts inside a radius of 8\farcs04 (quadratic sum of the PSF's $\sigma$ of the two instruments). When a single SPIRE source is associated with two (or more) different MIPS sources, we consider both associations. Both the  SPIRE and MIPS fluxes are proportional to the total IR luminosity  \citep[L$_{\mathrm{IR}}$, e.g. ][]{2001ApJ...556..562C, 2011A&A...533A.119E}. This is due to the same thermal origin of the radiation emitted in these bands. Given this assumption, when we find multiple MIPS counterparts for a single SPIRE  source, the original fluxes in the three SPIRE bands are divided among the MIPS counterparts proportionally to their 24 $\mu$m flux.
 This multiple association involves 429 MIPS sources, flagged in our catalog through a N\_MIPS\_SPIRE parameter greater than 1. SPIRE--MIPS associations outside the IRAC covered area are not included in our catalog. Using the MIPS--24 prior position, we found 7034 SPIRE counterparts for our IRAC sources.
 
\begin{figure}[!ht]
\begin{center}
\includegraphics[width=7.5cm]{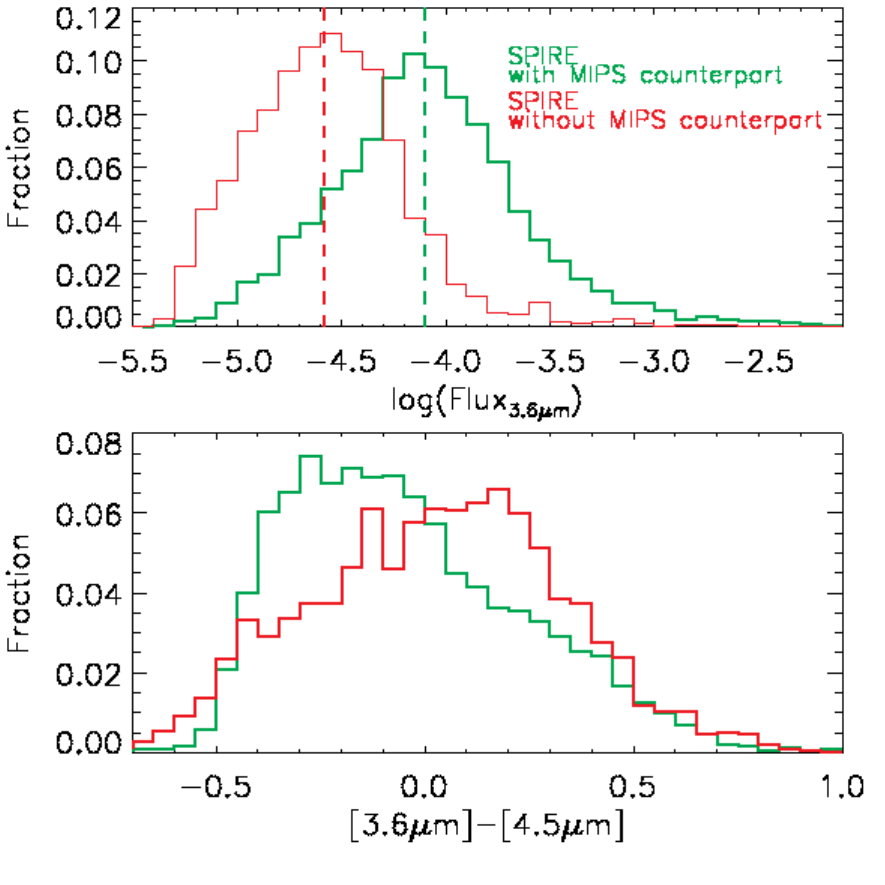}
\caption{\label{IMS_Fluxes_and_colors} MIPS--undetected SPIRE sources are likely located at higher redshift than MIPS detected sources. Normalized 3.6 \mic\ fluxes ({\bf top panel}) and [3.6 \mic]--[4.5 \mic] AB color distributions ({\bf bottom panel}) for SPIRE sources in our catalog. The SPIRE sources with and without MIPS counterparts are indicated in green and red respectively. The median of each distribution is represented with a dashed line in the same color code.}
\end{center}
\end{figure}

For the remaining SPIRE sources without a MIPS counterpart, we searched for a direct IRAC--SPIRE association.  We found 2413 SPIRE sources with associated IRAC counterparts. These are flagged in the catalog with N\_IRAC\_SPIRE$>$0 and N\_MIPS\_SPIRE$\leq$0.
The MIPS undetected SPIRE sources have  24 \mic\ flux  below the detection threshold in this band. This is because for SPIRE detected sources, the detection rate at 24 \mic\ strongly depends on the source redshift. This is due to the typical shape of the spectral energy distribution (SED) of far--IR detected sources\footnote{Some examples of typical SEDs for different type of galaxies can be found in e.g., \cite{2007ApJ...663...81P}.}, usually presenting a bump of emission centered at $\lambda^{\mathrm{dust}}_{\mathrm{R.F.}}\sim$100 \mic\ \citep{2012ApJ...759..139K}. This bump is due to the thermal emission of dust heated by optical--UV radiation produced by young stars inside star forming regions or by accretion disks of Active Galactic Nuclei (AGNs). At increasing redshifts, the SPIRE bands sample spectral regions closer to the redshifted peak.
Instead, being located in the opposite side of the thermal emission bump, the MIPS 24 \mic\ band samples a spectral region whose the emitted luminosity tends to be lower at higher redshifts. The combination of these effects is responsible for the lower MIPS detection rate among the SPIRE detected sources, at higher redshifts. This is demonstrated in Figure~\ref{IMS_Fluxes_and_colors}, where we show the distributions of 3.6\mic\ flux and [3.6\mic]--[4.5\mic] colors (i.e. $2.5\log(F_{4.5\mu m}/F_{3.6\mu m})$) for SPIRE sources with and without MIPS counterpart. 
The two IRAC bands sample the stellar 1.6\mic\ bump \citep{1999PASP..111..691S, 2002AJ....124.3050S} and the IRAC color is expected to change with redshift as the peak moves through the two filters. The color distributions in Figure~\ref{IMS_Fluxes_and_colors} are clearly different, indicating that  MIPS--undetected SPIRE sources are more commonly located at higher redshifts than the MIPS--detected SPIRE sources. This is also supported by the lower  median 3.6 \mic\ flux of the MIPS--undetected SPIRE sources (upper panel of Figure~\ref{IMS_Fluxes_and_colors}).

SPIRE sources  lacking an IRAC counterpart are generally not included in our catalog. However, we include in our catalog those 60 SPIRE sources  with a reliable MIPS counterpart but without any IRAC counterpart inside a distance corresponding to two times the IRAC--MIPS searching radius. We visually checked all sources in the IRAC 3.6 \mic\ image in order to exclude missed detections due to the presence of  nearby bright sources or border effects enhancing the noise and consequently the detection threshold.

The catalog includes a total of 9447 SPIRE sources with a SPIRE flux (in at least one band) above 3$\sigma$. The reliability of their association with the IRAC counterpart is discussed in the next Section.

\begin{deluxetable}{lcccccc}
\tabletypesize{\footnotesize}
\tablecolumns{7}
\tablewidth{0pc}
\tablecaption{Parameters Used in the Counterpart Identification Procedure\tablenotemark{a}.}
\tablehead{
\colhead{} & \colhead{Search}  & \multicolumn{2}{c}{Mean Distance \tablenotemark{b}} &\multicolumn{2}{c}{ Barycentre Position ($\Delta$RA, $\Delta$DEC) \tablenotemark{c}} & \colhead{} \\
\colhead{Band} & \colhead{Radius} & \colhead{Before R.}  & \colhead{After R.} & \colhead{Before R.} & \colhead{After R.}  & \colhead{$\Delta$N}  \\
\colhead{} & \colhead{(\farcs)} & \colhead{(\farcs)} & \colhead{(\farcs)} & \colhead{(\farcs)}  &  \colhead{(\farcs)} & \colhead{}   }
\startdata
MIPS                     & 2.60  & 1.01 & 0.87 &  0.098,  0.49   &  0.010, 0.071 & +140 (0.6\%)   \\
SPIRE \tablenotemark{d}  & 8.04  & 3.00 & 3.00 &  0.019, -0.045  &  0.0035, -0.0040 & -1 (0.01\%) \\ 
2MASS                    & 2.68  & 0.60 & 0.59 &  0.067, -0.067  &  0.0061, -0.0059 & +7 (0.005\%)\\
\enddata
\tablenotetext{a}{All the values are computed before the correction for average shift (Before R.) and after the correction (After R.).}
\tablenotetext{b}{Average distance, in arcseconds, between the IRAC 3.6 $\mu$m sources and the corresponding counterparts in the other bands.}
\tablenotetext{c}{Average difference ($\Delta$RA, $\Delta$dec) between the IRAC 3.6 $\mu$m sources and the corresponding counterparts in the other bands.}
\tablenotetext{d}{Distances refers to the MIPS 24 positions.}
\label{counterp_search_table}
\end{deluxetable}

\subsubsection{Counterpart Reliability}
\label{good_corr_sect}
\begin{figure}[!ht]
\begin{center}
\includegraphics[width=8cm]{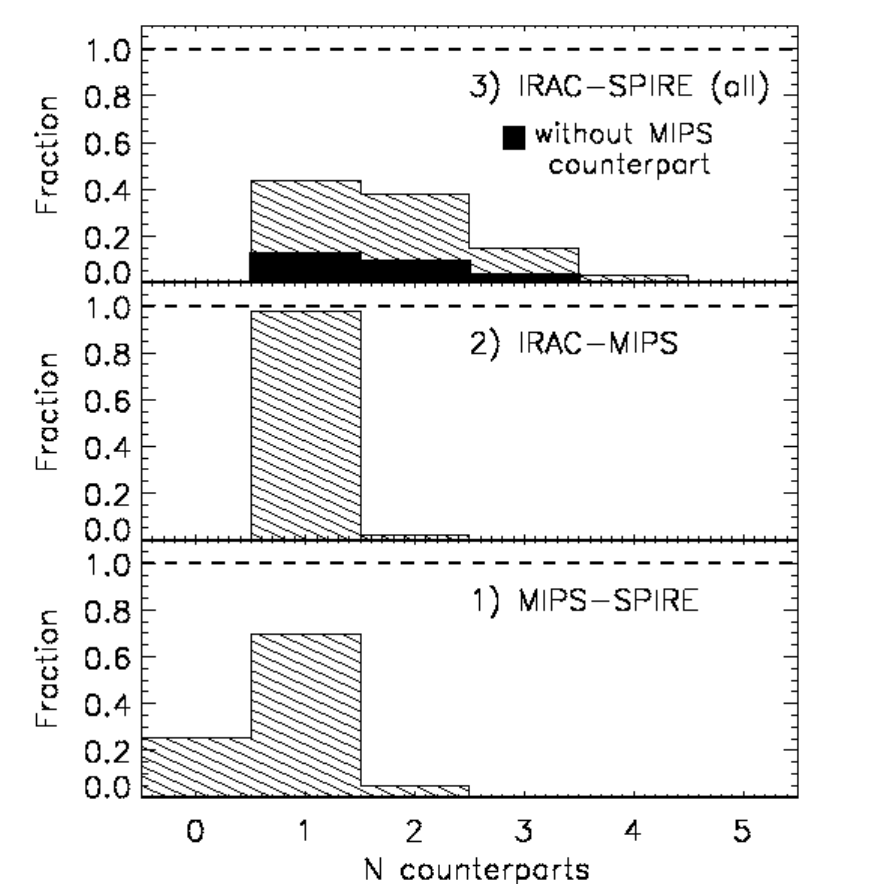}
\caption{\label{Ncnterpts}Fraction of sources with one or multiple counterparts inside the search radius for the SPIRE-MIPS, MIPS-IRAC, and SPIRE-IRAC positional  correlations. In the top panel, we show the distribution of the number of IRAC counterparts inside the SPIRE--IRAC search aperture for the full sample of SPIRE sources (hatched histogram).  The black filled histogram shows the distribution of the number of IRAC counterparts for SPIRE sources without a MIPS counterpart.  }
\end{center}
\end{figure}

The probability of having the right MIPS (IRAC)
counterpart associated to the SPIRE (MIPS) source depends on both the number of
galaxies found within the search area, as well as the distance to
the identified counterpart(s). %
In Figure~\ref{Ncnterpts} we show the fraction of MIPS and SPIRE sources with one or more IRAC or MIPS counterpart inside the searching radius. Only $\sim$2\% of our MIPS sources have more than one single IRAC counterpart inside the searching radius. In all these cases we consider the closest IRAC source as the only real counterpart. Instead, $\sim$6\% of our SPIRE--MIPS associations are made dividing the SPIRE fluxes among the multiple MIPS counterparts, as explained in Section~\ref{subsec:SPIRE}. In the direct SPIRE--IRAC associations we considered the closest IRAC sources as the only real counterparts, but in this case, only $\sim$48\% of our SPIRE source have a single IRAC counterpart inside the searhing radius.
In order to assess the reliability of the matched counterparts, we develop a parameter, $P$, that accounts for both the counterpart distance and the number of sources found in the searching radii.

As mentioned earlier, we define the search area by the radius $r_s$
equal to the the quadratic sum of the PSF's $\sigma$ of the two
instruments involved (e.g., r$_{s}=\sqrt{\sigma_{\mathrm{IRAC}}^2
  +\sigma_{\mathrm{MIPS}}^2}$, for IRAC--MIPS correlation).  We
consider a normalized 2D Gaussian function with $\sigma=r_s$; the counterpart distance $d_i$ is always smaller then $r_s$. 
We then compute the quantity
$A_i$ as the probability of the galaxy being at a distance greater
than $d_i$, for the given Gaussian function. For a single counterpart,
centered on the coordinates of the starting objects, $A_i=1$; in
general, $A_i$ decreases as the distance of the counterpart increases.
We then account for the presence of multiple counterparts (at
different $d_i$), by defining the parameter $P$ as follows:

\begin{equation}
\mathrm{P}=\mathrm{A}_{1} \frac{\mathrm{A}_{1}}{\sum_{i}{ \mathrm{A}_{i}} },
\end{equation}

\noindent
where $A_1$ is defined as $A_i$ corresponding to the closest
counterpart.  For multiple counterparts the factor $A_1/\sum{A_i}$ is
smaller than one, and decreases with the number of counterparts.

\begin{figure}[!ht]
\begin{center} 
\includegraphics[width=8cm]{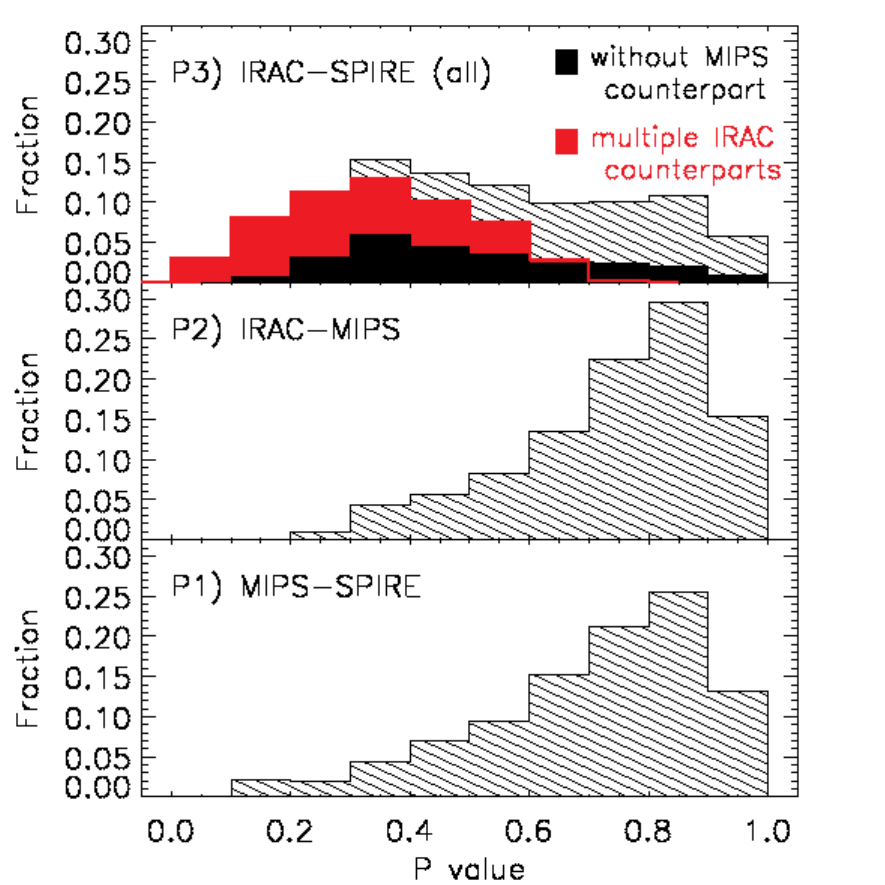}
\caption{\label{indicator_P}Correlation reliability indicator $-P-$ computed for the SPIRE-MIPS (P1), MIPS-IRAC (P2), and SPIRE-IRAC (P3) positional correlations. The latter was computed for all SPIRE sources in our catalog, even when the correlation with IRAC is found through the MIPS position. The black filled distribution in the top panel shows SPIRE sources without a MIPS counterpart. The red filled distribution is computed for SPIRE sources with multiple IRAC counterparts inside the IRAC-SPIRE search radius.} \end{center}
\end{figure}

We computed the values of $P$ for the SPIRE-MIPS ($P1$), MIPS-IRAC ($P2$), and SPIRE-IRAC ($P3$) correlations. The distributions of the $P$ values are represented in Figure~\ref{indicator_P}. The direct SPIRE-IRAC correlation is studied for the whole SPIRE sample in our catalog, considering also the sources for which we found a correlation through the MIPS position.

The MIPS-IRAC and SPIRE-MIPS association reliability is usually high, as demonstrated by the distribution of $P1$ and $P2$. As a consequence, the association of the SPIRE sources to the IRAC counterparts through the MIPS position is still reliable, even if the $P3$ distribution is not as narrow as the $P1$ and $P2$ ones.

Using a Monte Carlo simulation, we assessed the reliability of those associations in which multiple MIPS counterparts are associated to individual SPIRE sources (see Section~\ref{subsec:SPIRE}). The same simulation also allows us to determine the reliability of the direct SPIRE--IRAC associations.
The simulation is performed by first randomly shifting the position of the SPIRE sources to 10, 20, 30, and 50 times the search radius from the original position and then looking for potential IRAC counterparts. We found a detection rate of approximately 49.7\%$\pm0.1$\%

In our catalog, among the SPIRE counterparts found through a MIPS prior position, the fraction of SPIRE sources having multiple IRAC potential associations  is 57.6\%. SPIRE sources with MIPS counterparts are generally associated to high flux IRAC counterparts: only  4.0\% of them have an IRAC flux below the 90\% completeness limit (see Table~\ref{completenessI1_table}). Moreover, multiple IRAC counterparts in a MIPS--IRAC search radius are very rare. This means that pure-geometrical MIPS--IRAC associations are highly improbable. Therefore, we can safely assume that for almost all our SPIRE counterparts found through MIPS prior positions, there is a detected IRAC real counterpart. Given these assumptions and the results of our simulation, we expect that 49.7\% of the SPIRE sources have, beside the real IRAC counterpart, an additional purely geometrical association. Since we measure a real multiple association rate of 57.6\%, the additional 7.9\% of these SPIRE sources must have real multiple IRAC components. Indeed, in our catalog, $\sim$6.1\% of  the SPIRE sample have multiple MIPS counterparts, and all of them have one IRAC counterpart inside the MIPS--IRAC search radius.
The $\sim$2\% difference confirms the reliability of the multiple SPIRE--MIPS associations discussed in Section~\ref{subsec:SPIRE}.

As can be observed in the top panel of Figure~\ref{IMS_Fluxes_and_colors}, the 3.6 \mic\ flux for the SPIRE sources is generally lower when they are not detected at 24 $\mu$m. As explained in section \ref{subsec:SPIRE}, the lower IRAC flux of these MIPS undetected sources can be explained with the higher median redshift of the these sources. We find that $\sim$24\% of this sample are below the 90\% completeness level, resulting in an overall completeness of $\sim$85\%. Therefore, the probability of detecting the real counterpart inside the search radius is $P_{\mathrm{det}}^{\mathrm{real}}$=0.85 while, from our simulation, the probability to find a purely geometrical IRAC counterpart is $P_{\mathrm{det}}^{\mathrm{geom}}$=0.497.  We define the following products: 

\begin{eqnarray}
P_{1}=P_{\mathrm{det}}^{\mathrm{geom}} (1-P_{\mathrm{det}}^{\mathrm{real}}),\\
P_{2}=P_{\mathrm{det}}^{\mathrm{geom}} P_{\mathrm{det}}^{\mathrm{real}},\\
P_{3}=(1-P_{\mathrm{det}}^{\mathrm{geom}}) (1-P_{\mathrm{det}}^{\mathrm{real}}),\\
P_{4}=(1-P_{\mathrm{det}}^{\mathrm{geom}}) P_{\mathrm{det}}^{\mathrm{real}}. 
\end{eqnarray}

Each SPIRE source can either have a possible counterpart (with probability $P_{\mathrm{det}}=P_{1}+P_{2}+P_{4}$) or not ($P_{\mathrm{no-det}}=P_{3}$), with $P_{1}+P_{2}+P_{3}+P_{4}=1$.
All SPIRE sources in our catalog have one IRAC association. Assuming that the SPIRE position is always closer to the real IRAC counterpart than to the nearest geometrical association, their reliability can be estimated as:
\begin{eqnarray}
P^{\mathrm{real}}_{\mathrm{det}}(\mathrm{cat})=\frac{P_{2}+P_{4}}{P_{1}+P_{2}+P_{4}}\sim 0.92
\end{eqnarray}

In reality, if a purely geometrical IRAC counterpart is present, it can be closer to the SPIRE position than the real counterpart. This is more probable when the IRAC--SPIRE distance is higher (and consequently the $P3$ parameter lower).
Moreover, as stated before, about 6--8\% of the SPIRE sources have more than a single real counterpart, further reducing the reliability for this sample. For all  SPIRE sources in our catalog we report the number of potential IRAC counterparts in the search radius (N\_IRAC\_SPIRE) and the parameter $P3$.


\subsection{Optical R$_{c}$--Band}
\label{WFI_R_SECT}

A central area of approximately one square degree was observed at the MPG/ESO 2.2 m telescope at La Silla with the Wide Field Imager (WFI) in October 2010 (P.I.: T. Takeuchi). The 8 CCDs of the WFI camera cover a total area on the sky of 34$'\times 33'$, with a pixel scale of 0\farcs24. Four pointings with the R$_c$ broad band filter ($\lambda_{c}=6517.25$\AA) were obtained, covering a total area of 1.13 deg$^{2}$. Each pointing was observed with multiple exposures dithered to optimally remove the gaps between different CCDs and other CCD defects. The total exposure time varied between 2.2 and 1 hour.  

The data were reduced with standard IRAF routines included in the  NOAO mosaic software MSCRED. A Super-Sky Flat-Field (SSFF) correction was applied by dividing all science frames for the average of the non-aligned and source--subtracted science exposures.
The final R$_{c}$  mosaic was created combining the images on the four pointings.
In the final mosaic, the full width half maximum (FWHM) of the PSF was 1\farcs0.

Sources were extracted from the final mosaic using the \emph{SExtractor} software.  We considered only sources with five connected pixels above a threshold of 
1.0$\sigma$ of the local background
For each detected object, we recorded in the catalog the total AUTO flux. 
The photometric calibration of these data is obtained through the comparison with \citet[][BC03]{2003MNRAS.344.1000B} template SEDs fitted to a large set of optical data. We used a set of measures obtained in the SIMES field for a selected sample of sources, in 13 different filters covering the spectral range between the u and the IRAC 4.5 \mic\ bands (I. Baronchelli et al. in preparation). Knowing the spectroscopic redshift of the selected sources \citep{2011MNRAS.416.1862S}, we used a $\chi^{2}$ minimization technique \cite[i.e. hyperzmass,][]{2000A&A...363..476B} to find the best fitting SED among the BC03 template SEDs. After comparing the extracted R$_{c}$ flux with the expected flux obtained from the convolution of the WFI--R$_{c}$ filter response with the best fitting SED, we used the average difference to calibrate the extracted R$_{c}$ fluxes.

We computed the depth  of the optical mosaic as follows. We measured the flux inside randomly distributed 1\farcs9
 diameter apertures and we fitted a Gaussian function to the symmetrized distribution of flux values. The 3$\sigma$ flux limit of the R$_c$ image is 0.53$\mu$Jy.

In order to combine the optical and IR data, we searched the IRAC catalog for the closest counterpart to each $R-$detected source, using a search radius of 0\farcs82. The precise technique adopted is described in I. Baronchelli et al. (in preparation). We found an optical counterpart for $\sim$55\% of the IRAC-detected sources in our catalog and which are covered in the R$_{c}$ band

\subsection{Galaxy/Star Separation}
\label{gal_star_separation}

We perform the galaxy/star separation only in the central square degree area covered by the optical data. Our separation criteria, described in detail below, are based on a combination of diagnostics using optical, 3.6 and 24 \mic\ fluxes, as well as the surface brightness profile of each source from the optical data. Our selection criteria were calibrated using the stellar spectral models of \cite{1993KurCD..13.....K} and a representative set of galaxy SED templates from \cite{2007ApJ...663...81P}, including ellipticals, spiral galaxies, starburst galaxies and QSO templates.

\begin{figure}[!ht]
\begin{center}
\includegraphics[width=7.5cm]{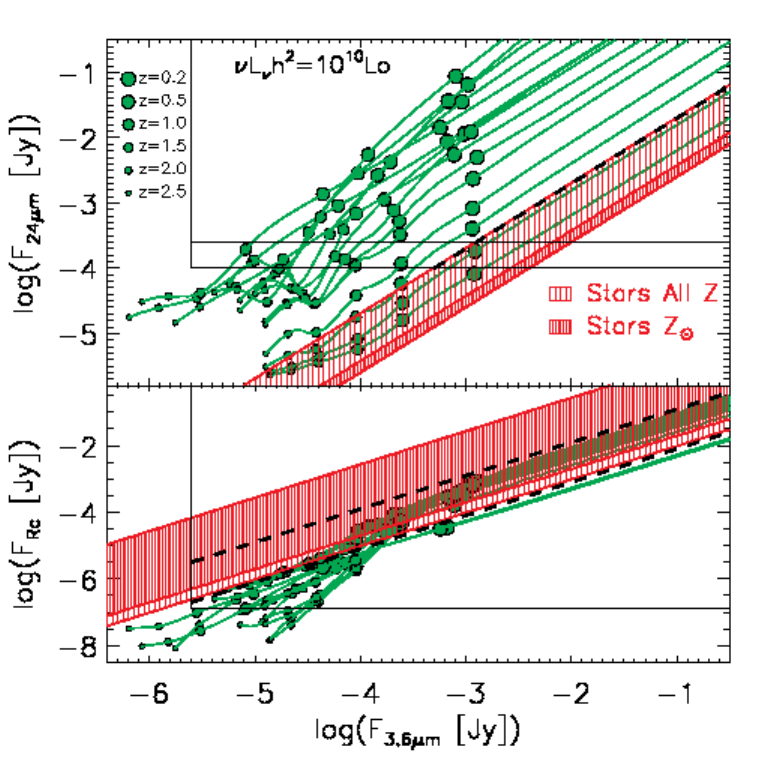}
\caption{\label{gal_stars_track} IRAC 3.6, MIPS 24 \mic,\ and R$_{c}$ fluxes for a library of templates including both stars \citep[red, using the stellar models of][]{1993KurCD..13.....K} and galaxies (green, in the range 0$<$z$<$2.5 and normalized to $\nu$L$_{\nu}h^{2}$=10$^{10}$L$_{\odot}$ at 3.6 $\mu$m).   The selection thresholds used are represented with dashed black lines. The black boxes represent the areas covered by the plots of Figure~\ref{star_selection}. } 
\end{center}
\end{figure}

We considered galaxy models in the range 0$<$z$<$2.5 and with $\nu$L$_{\nu}h^{2}$=10$^{10}$L$_{\odot}$ at 3.6 $\mu$m,  close to the characteristic $L^*$ luminosity  \citep[e.g. ][]{2006A&A...453..397F, 2006MNRAS.370.1159B}. We used the \cite{1993KurCD..13.....K} stellar models, with abundances relative to solar ranging from $\log(Z)=1.0$ to $\log(Z)=-5.0$. 
In Figure~\ref{gal_stars_track}, the differences between the areas occupied by the solar metallicity models and by the models with all the possible metallicities are shown.  For each modeled star or galaxy, we computed the expected fluxes in the 3.6 \mic, 24 $\mu$m, and R$_{c}$ bands.
In Figure~\ref{gal_stars_track}, galaxy tracks are shown in green (different sizes correspond to different redshifts). Regions in the diagnostic diagrams occupied by stars are shown as shaded red band.

\begin{figure}[!ht]
\begin{center}
\includegraphics[width=7.5cm]{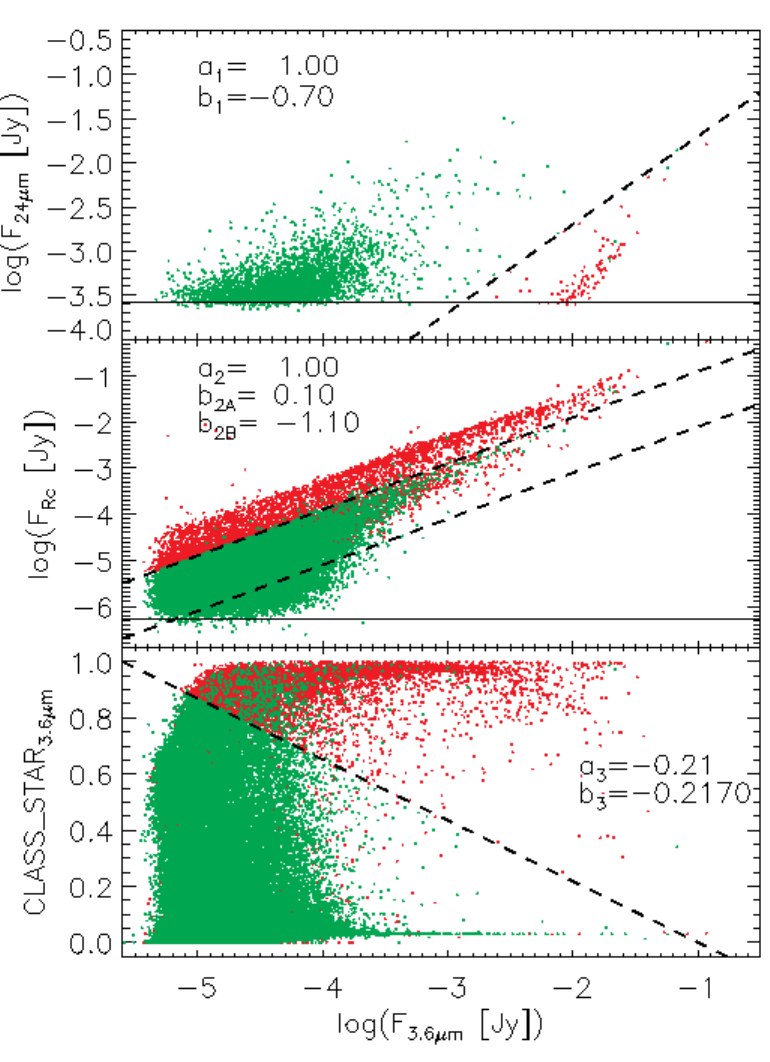}
\caption{\label{star_selection} Galaxy/star separation. Same as Figure~\ref{gal_stars_track}, but plotting the measurement for the objects detected in the SIMES field. In each panel, green points show sources that have been classified as galaxies by one of our criteria (see text for details).}
\end{center}
\end{figure}

The bulk of the stellar emission, for a galaxy, is located at $\lambda\sim$ 1.6 $\mu$m \citep{1999PASP..111..691S, 2002AJ....124.3050S}. Stars are in general fainter at longer wavelengths, especially at the SPIRE wavelengths, where the galaxy spectra is dominated by the dust thermal emission. For this reason, among the SPIRE sources, the probability of detecting a star is negligible if compared to that of detecting a galaxy. Therefore, we classify as galaxy any object detected in one of the SPIRE bands. This assumption is confirmed by a visual inspection of the SPIRE sources on the IRAC images, where the bright 3.6 $\mu$m saturated stars are not SPIRE detected.

 Then, following our diagnostics, we identify stars in the 3.6 \mic\ versus MIPS--24 plane (top panels of Figures~\ref{gal_stars_track} and \ref{star_selection}). 
All  sources with $\log (F_{24}[\mathrm{Jy}])< \log (F_{3.6\mu m})-0.7$ are classified as stars.
Because of the bright MIPS--24 flux limit ($\log F_{24}[\mathrm{Jy}] \sim$ --3.6), this selection misses faint
IRAC--detected stars. We thus implement two additional constraints,
based on the R$_{c}$-3.6 \mic\ color (bottom panel of Figure~\ref{gal_stars_track} and middle panel of Figure~\ref{star_selection}) and the \emph{SExtractor} CLASS\_STAR\footnote{CLASS\_STAR $=0$ for galaxies, $=1$ for stars. This \emph{SExtractor} output parameter quantifies the similarities between    a source surface brightness profile and the profile of a point--like source.} parameter measured in the 3.6 \mic\ image (bottom panel in Figure~\ref{star_selection}). The SPIRE and 24 \mic\ undetected sources are classified as stars when 
$\log (F_{R_{c}}[\mathrm{Jy}])> \log (F_{3.6\mu m})+0.1$,  and as galaxies when 
$\log (F_{R_{c}}[\mathrm{Jy}])< \log (F_{3.6\mu m})-1.1$. Between these two limits, or when no SPIRE, MIPS or R$_{c}$ counterparts are found, we rely on the combination of the 3.6 \mic\ flux and the CLASS\_STAR parameter to identify stars in our sample.

The reliability of the \emph{SExtractor} CLASS\_STAR parameter  worsens at the faintest and  brightest IRAC fluxes. At low fluxes (i.e. 
$F_{3.6\mu m}\lesssim 10^{-4}$ Jy)
, the shape of a galaxy looks similar to that of a point--like source, while the PSF wings of the brightest objects (i.e. 
$F_{3.6\mu m}\gtrsim 10^{-2.0}$ Jy) can be incorrectly interpreted as due to an extended profile by \emph{SExtractor}.
For these reasons, we introduce a flux dependent CLASS\_STAR threshold: a source is identified as a star if 
$ \mathrm{CLASS\_STAR} > -0.21 \log (F_{3.6\mu m}[\mathrm{Jy}])-0.217 $.
We visually checked the correct identifications of all bright sources  
($F_{3.6\mu m}>10^{-3.2}$ Jy) using the R$_{c}$ image, correcting our counts for saturated stars wrongly identified as galaxies.

In order to assess the reliability of our diagnostic method, we compared our stellar number counts with those expected from a Milky Way model of stellar distribution. To compute the simulated stellar number counts we used the population synthesis code \emph{TRILEGAL}\footnote{More information on this code can be found at: http://stev.oapd.inaf.it/cgi-bin/trilegal} \citep{2005A&A...436..895G}, considering the position of the SIMES field. 
The result of this comparison is visible in Figure~\ref{counts_IRAC1}, where the simulated stellar number counts are represented with a dashed line while the counts of stars identified through our diagnostic method are reported using a dotted line. At all fluxes, we observe a good agreement between observed and simulated stellar counts, confirming the reliability of our method.

\section{RESULTS}
\label{sec:results}
The IRAC--based multiwavelength photometric catalog together with the IRAC 3.6 and 4.5 \mic\ mosaics are released through the NASA/IPAC Infrared Science Archive (IRSA). Table~\ref{catalog_cont_table} describes all the columns in our photometric catalog. 
In the following sections we use this catalog to measure galaxy integral number counts (Section~\ref{Counts}) and to search for $z\gtrsim 1.3$ galaxy clusters (Section~\ref{sec:clusters}). These are preliminary results,  illustrative of those that will be allowed by the survey. We anticipate that future papers will improve the analysis once the deep optical Dark Energy Survey \citep[DES, ][]{2005IJMPA..20.3121F} and CTIO-r (L. Barrufet et al. in preparation) data become available over the full survey area.

\subsection{Cumulative Number Counts}
\label{Counts}
We derive the galaxy--only and galaxy$+$stars (\emph{total}) number counts from the 3.6 \mic\ map, and compare them with the   results of 
\citet[][FA04]{2004ApJS..154...39F}, \citet[][FR06]{2006A&A...453..397F}, and \citet[][AS13]{2013ApJ...769...80A}. We computed the \emph{total} counts in the entire SIMES field and the \emph{galaxies}--only counts in the central square degree, where we use the optical R$_{c}$ band for the galaxy/star separation. All the number counts are corrected for incompleteness, with the values presented in Table~\ref{completenessI1_table}. We report the completeness--corrected SIMES 3.6 \mic\ integral number counts (both for galaxies$+$stars, and galaxies only), with the associated uncertainties, in Table \ref{counts_I1_table}. 

\begin{figure}[!ht]
\begin{center}
\includegraphics[width=7.5cm]{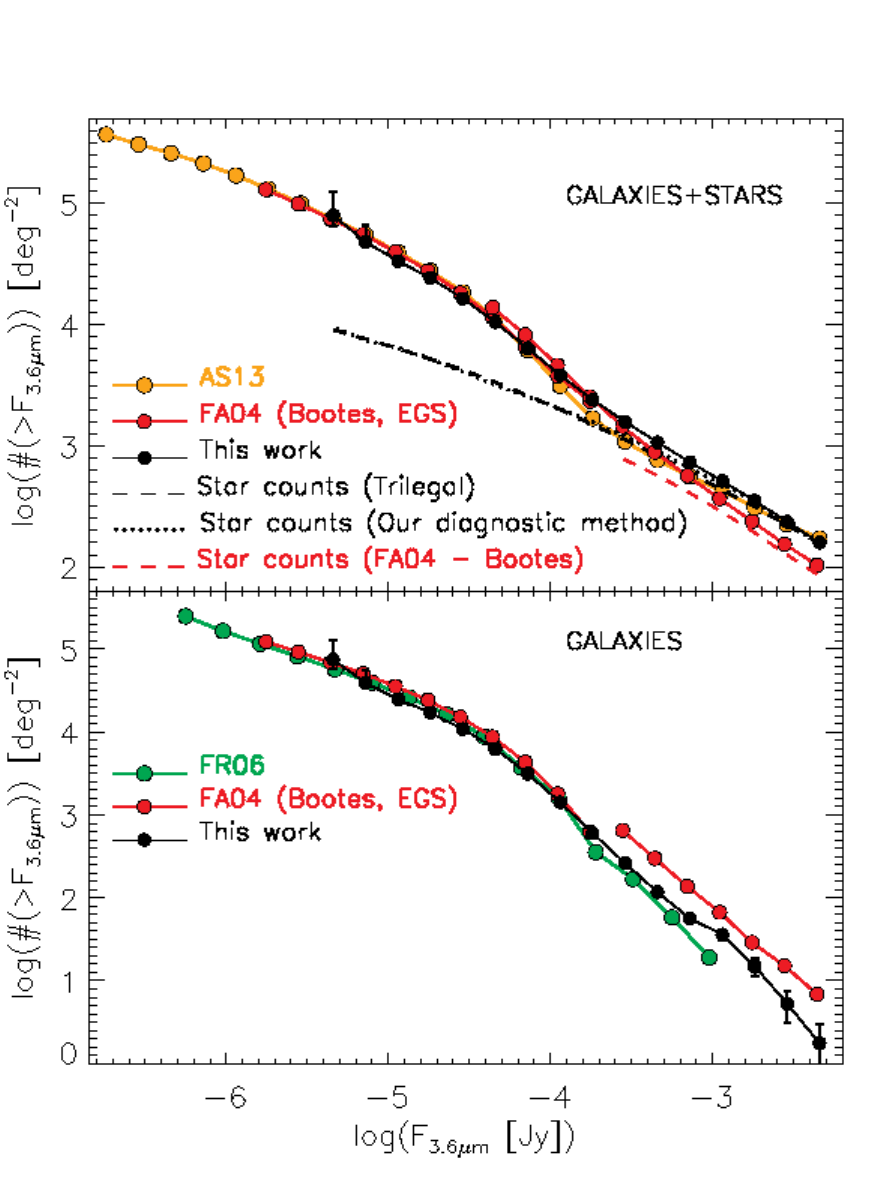}
\caption{\label{counts_IRAC1} {\bf Top panel} \emph{Total} (galaxy$+$stars) completeness corrected integral number counts at 3.6 \mic\ from the SIMES survey (black symbols and line). Literature results are also shown in the Figure as indicated by the legend, with AS13=\cite{2013ApJ...769...80A}, FA04=\cite{2004ApJS..154...39F}, FR06=\cite{2006A&A...453..397F}. The observed and simulated (using \emph{TRILEGAL} code) star counts computed in the SIMES field are represented with dotted and dashed black lines. FA04 star counts in the Bo\"{o}tes field are represented with a dashed red line. The FA04 counts are reported here for both the EGS (faint end) and the Bo\"{o}tes (bright end) field. For the Bo\"{o}tes field, source counts are provided fainter than stars and galaxies could be reliably separated using the method described in FA04. Above $F_{3.6 \mu m} \sim 10^{-3.7}$ Jy, the \emph{total} counts of AS13  are fully reproduced by  stars alone.}
\end{center}
\end{figure}

The comparison between the cumulative SIMES number counts below 10$^{-2.35}$ Jy and those presented in \cite{2004ApJS..154...39F}, \cite{2006A&A...453..397F} and \cite{2013ApJ...769...80A} are shown in
Figure~\ref{counts_IRAC1}. Out of the three fields presented in FA04,
we only compare to their low and intermediate depth fields (more similar to the data presented here), i.e., $\sim$3°$\times$3° area in the the Bo\"{o}tes field and
0\fdg17$\times$2\fdg0 area in the EGS field respectively. The FA04 \emph{galaxy} and \emph{total} number counts in the EGS field are presented in differential form, starting at the highest flux of 10$^{-3.76}$ Jy. 
To convert them from differential to cumulative, we use our cumulative \emph{galaxy} number counts at 10$^{-3.76}$ Jy as starting point. For the FA04 counts in the Bo\"{o}tes field we do not use any starting point, since they are reported through high fluxes (i.e. 10$^{-1.76}$ Jy) where counts assume fractional values.
FR06 presents cumulative number counts for galaxy only and no starting point is needed. On the other hand, AS13 presents \emph{total} differential counts. As for the FA04 EGS counts, we used the SIMES cumulative counts at the flux of the brightest AS13 bin as staring point.

\begin{deluxetable}{cccccccccc}
\tabletypesize{\footnotesize}
\tablecolumns{8}
\tablewidth{0pc}
\tablecaption{Raw and Completeness Corrected Integral Number Counts at 3.6 $\mu\mathrm{m}$}
\tablehead{
\colhead{Flux} & \multicolumn{2}{c}{Raw Counts ($>S_{3.6}$)}  & \colhead{Corr.} & \multicolumn{6}{c}{Corrected Counts ($>S_{3.6}$)} \\
\colhead{(log[Jy])} & \colhead{N$_{\mathrm{GAL}}$} & \colhead{N$_{\mathrm{TOT}}$} & \colhead{Fact.}  & \colhead{N$_{\mathrm{GAL}}$} & \colhead{N$_{\mathrm{GAL}}^{\mathrm{Inf}}$} & \colhead{N$_{\mathrm{GAL}}^{\mathrm{Sup}}$} & \colhead{N$_{\mathrm{TOT}}$} & \colhead{N$_{\mathrm{TOT}}^{\mathrm{Inf}}$} & \colhead{N$_{\mathrm{TOT}}^{\mathrm{Sup}}$} \\
        &      \colhead{(deg$^{-2}$)}           &  \colhead{(deg$^{-2}$)}                 &       &  \colhead{(deg$^{-2}$)}&  \colhead{(deg$^{-2}$)} &  \colhead{(deg$^{-2}$)}&  \colhead{(deg$^{-2}$)}& \colhead{(deg$^{-2}$)}& \colhead{(deg$^{-2}$)} }
\startdata
 -2.34 &     2 $\pm$   1 &   159 $\pm$  13 &  1.0  &       2 & 1 & 3      &    159&    154 &     163   \\
 -2.54 &     5 $\pm$   2 &   236 $\pm$  15 &  1.0  &       5 & 3 & 7      &    236&    230 &     241  \\
 -2.74 &    15 $\pm$   4 &   352 $\pm$  19 &  1.0  &      15 & 11 & 19     &    352&    346 &     359  \\
 -2.94 &    36 $\pm$   6 &   511 $\pm$  23 &  1.0  &      36 & 30 & 42     &    511&    503 &     519   \\
 -3.14 &    56 $\pm$   8 &   735 $\pm$  27 &  1.0  &      56 & 49 & 63    &    735&    725 &     745    \\
 -3.34 &   117 $\pm$  11 &  1069 $\pm$  33 &  1.0  &     117 & 107 & 127    &   1069&   1057 &    1081  \\
 -3.54 &   263 $\pm$  16 &  1583 $\pm$  40 &  1.0  &     263 & 247 & 278    &   1583&   1568 &    1597  \\
 -3.74 &   604 $\pm$  25 &  2388 $\pm$  49 & 0.997 &    603 & 580 & 627   &   2395&   2377 &    2414   \\
 -3.94 &  1416 $\pm$  38 &  3803 $\pm$  62 & 0.990 &   1421 & 1385 & 1457   &   3841&   3817 &    3865  \\
 -4.14 &  3116 $\pm$  56 &  6250 $\pm$  79 & 0.981 &   3150 & 3096 & 3204   &   6373&   6341 &    6405    \\
 -4.34 &  6151 $\pm$  78 & 10187 $\pm$ 101 & 0.972 &   6262 & 6188 & 6338  &  10481&  10443 &   10522  \\
 -4.54 & 10646 $\pm$ 103 & 15832 $\pm$ 126 & 0.963 &  10904 & 10802 & 11010  &  16434&  16376 &   16503  \\
 -4.74 & 16629 $\pm$ 129 & 23149 $\pm$ 152 & 0.954 &  17143 & 17012 & 17292  &  24274&  24194 &   24391   \\
 -4.94 & 23892 $\pm$ 155 & 31579 $\pm$ 178 & 0.941 &  24859 & 24679 & 25968  &  33543&  33407 &   34820   \\
 -5.14 & 32069 $\pm$ 179 & 39748 $\pm$ 199 & 0.827 &  38908 & 34570 & 56867  &  48091&  43760 &   66032   \\
 -5.34 & 36886 $\pm$ 192 & 43923 $\pm$ 210 & 0.549 &  74959 & 57300 & 126954 &  79997&  64693 &  125058   \\
\enddata
\tablenotetext{a}{For the raw counts, the uncertainty is the Poissonian error, while for the completeness corrected counts, we also consider the asymmetrical uncertainty on the estimated completeness curve.}
\label{counts_I1_table}
\end{deluxetable}

At 3.6 \mic\ fluxes fainter then $\sim 10^{-4}$Jy, our cumulative \emph{galaxy} number counts, shown in the bottom panel of Figure~\ref{counts_IRAC1}, agree well with the three reference surveys.
Above $\sim 10^{-4}$Jy differences are observed among all the works. Our galaxy counts  fall in  between those of FA04 for the Bo\"{o}tes field and the FR06 counts. The scatter at bright fluxes  is likely ascribable to two main reasons. First, the field to field variation combined with the bias against bright sources affecting deep field small areas. The deep GOODS-South area \citep{2003mglh.conf..324D} analyzed in FR06, among other reasons, was selected for being far from bright sources.
A second likely explanation is the uncertainty in the galaxy/star separation. In particular, we note that the difference between our galaxy counts in the brightest bin and the FA04 counts at the same flux level can be fully explained with a $<$5\% uncertainty in our corresponding stellar counts. 

The star counts in the SIMES area agree well with the counts simulated using the \emph{TRILEGAL} software \citep{2005A&A...436..895G}. For the \emph{TRILEGAL} simulation, based on a statistical description of the stellar distribution in the Milky Way, we set the same coordinates of the SIMES field in order to obtain comparable results. Instead, the FA04 stellar counts refer to a different area of the sky (Bo\"{o}tes), where stellar counts are expected to be lower, because of the higher galactic latitude. This is indeed observed in FA04 (upper panel of Figure~\ref{counts_IRAC1}) and it is confirmed by the results of a second \emph{TRILEGAL} simulation that we performed in an area centered in the (Bo\"{o}tes) region. In the brightest flux bin, simulated stellar counts resulted $\sim$20\% lower than FA04 real counts. This indicates that the different galaxy counts in the SIMES and Bo\"{o}tes fields are unlikely due to an underestimation of FA04 stellar counts.
Similar differences in stellar counts, when comparing different areas of the sky, can be observed e.g. in Papovich C. et al. 2015, ApJ, submitted.

\subsection{Selection of Clusters at intermediate Redshifts }
\label{sec:clusters}
Galaxy clusters -- through their space density and  evolution with cosmic time -- provide crucial information on the physical processes involved in cosmic structure formation. Representing the most extreme density environments, they provide galaxy samples with near coeval formation histories, and
are thus ideal laboratories in which to investigate the interplay between galaxy evolution and
environment, including the relative importance of triggering/quenching of star-formation and
AGN activity on galaxy assembly. \citet{2010ApJ...716.1503P}  extended the search for galaxy clusters to $z > 1.5$ by selecting galaxy-cluster candidates from the SWIRE survey solely as over-densities of galaxies with red IRAC colors, satisfying [3.6]$-$[4.5]$> -0.2$ magnitudes \citep[see also][for a recent application of this technique]{2014ApJ...797..109R}. The idea behind the method is simple, and is based on the 1.6 \mic\ stellar peak progressively moving out of the 3.6 \mic\ and entering the 4.5 \mic\ filter as redshift increases above $z\sim0.7$. 

In order to identify overdensities of red galaxies we proceeded in a similar way as \citet{2014ApJ...797..109R}. Briefly, before searching for spatial over densities, we preselected only those galaxies which satisfied the following conditions: IRAC [3.6]$-$[4.5]$> -0.2$, $19.5<$[4.5]$<21.5$, and $S/N > 3$ and 5, at 3.6 and 4.5 \mic, respectively. Similar cuts have been effectively used by various programs \citep{2010ApJ...716.1503P, 2010A&A...522A..58G, 2013A&A...559A...2G, 2012ApJ...759L..23G,2013ApJ...767...39M,2014ApJ...797..109R}. In the central square degree where our WFI--R$_{c}$ data are available, we also require selected galaxies to have F$_{\mathrm{R}_{c}}<14.5$ $\mu$Jy (this condition helps to broadly reject contaminants at $z<0.3$). For each galaxy $j$ in the selected sample, we then computed the quantity $\Delta N_{CC} = N_{1'} - N_{5'-8'}$, i.e., the difference between the number of red galaxies within 1\farcm0 from the $j^{th}$ galaxy ($N_1$) and the number of red galaxies in the background, that we computed inside an annulus of radius 5\farcm0 to 8\farcm0, normalized  to a circular area of 1\farcm0 radius ($N_{5'-8'}$).  All counts are corrected for incompleteness  using the results of the simulation discussed in Section~\ref{sec:completeness}, and computing the number of galaxies within a given distance from galaxy $j$, as: $N_j=\sum_i^N \frac{1}{C_i(3.6)}$, where the sum is over all N galaxies within 1\farcm0 from galaxy $j$, and  $C_i$ is the completeness corresponding to the 3.6 \mic\ flux of galaxy $i$.   

\begin{figure}[ht!]
\begin{center}
\includegraphics[width=7.5cm]{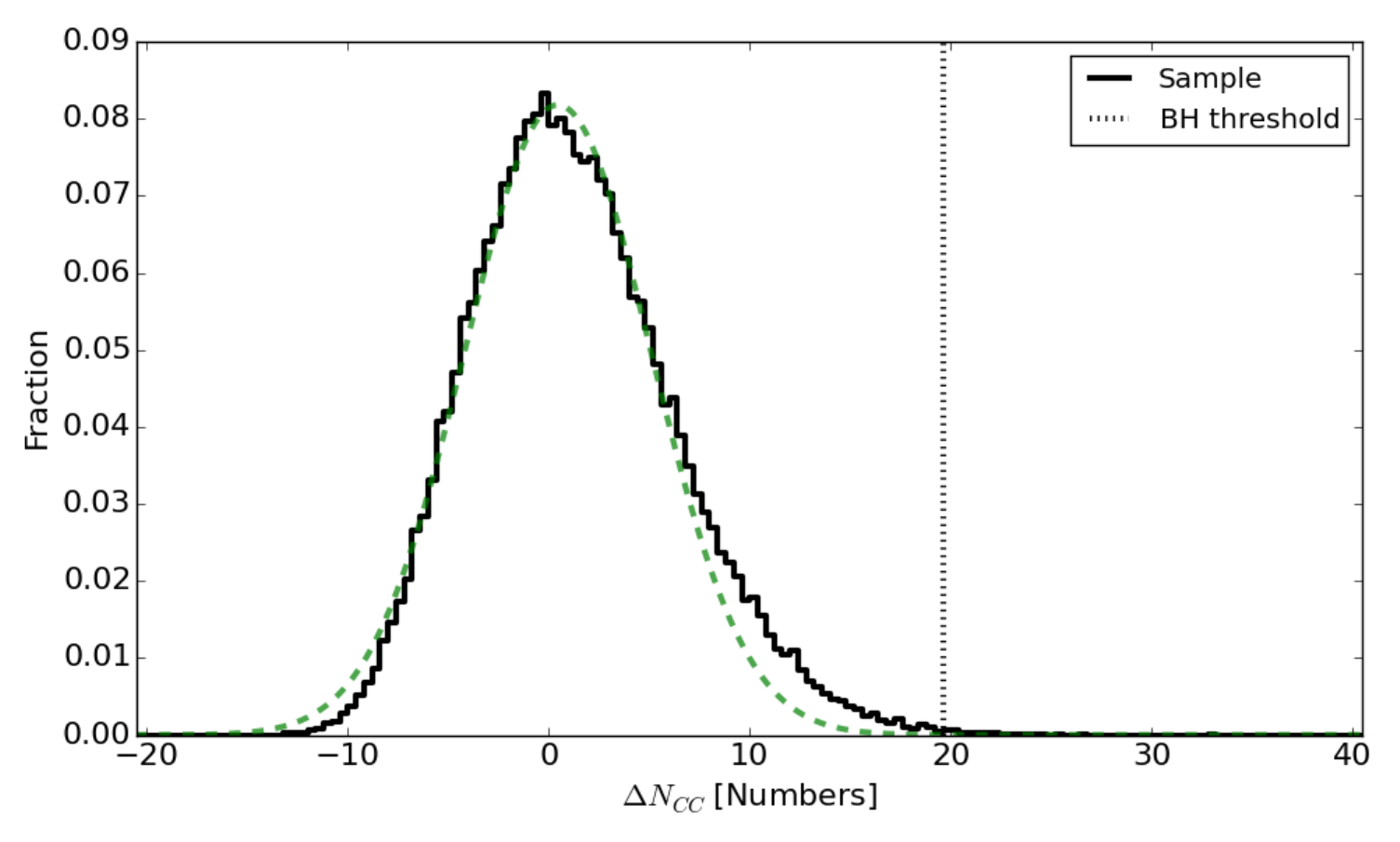}
\includegraphics[width=7.5cm]{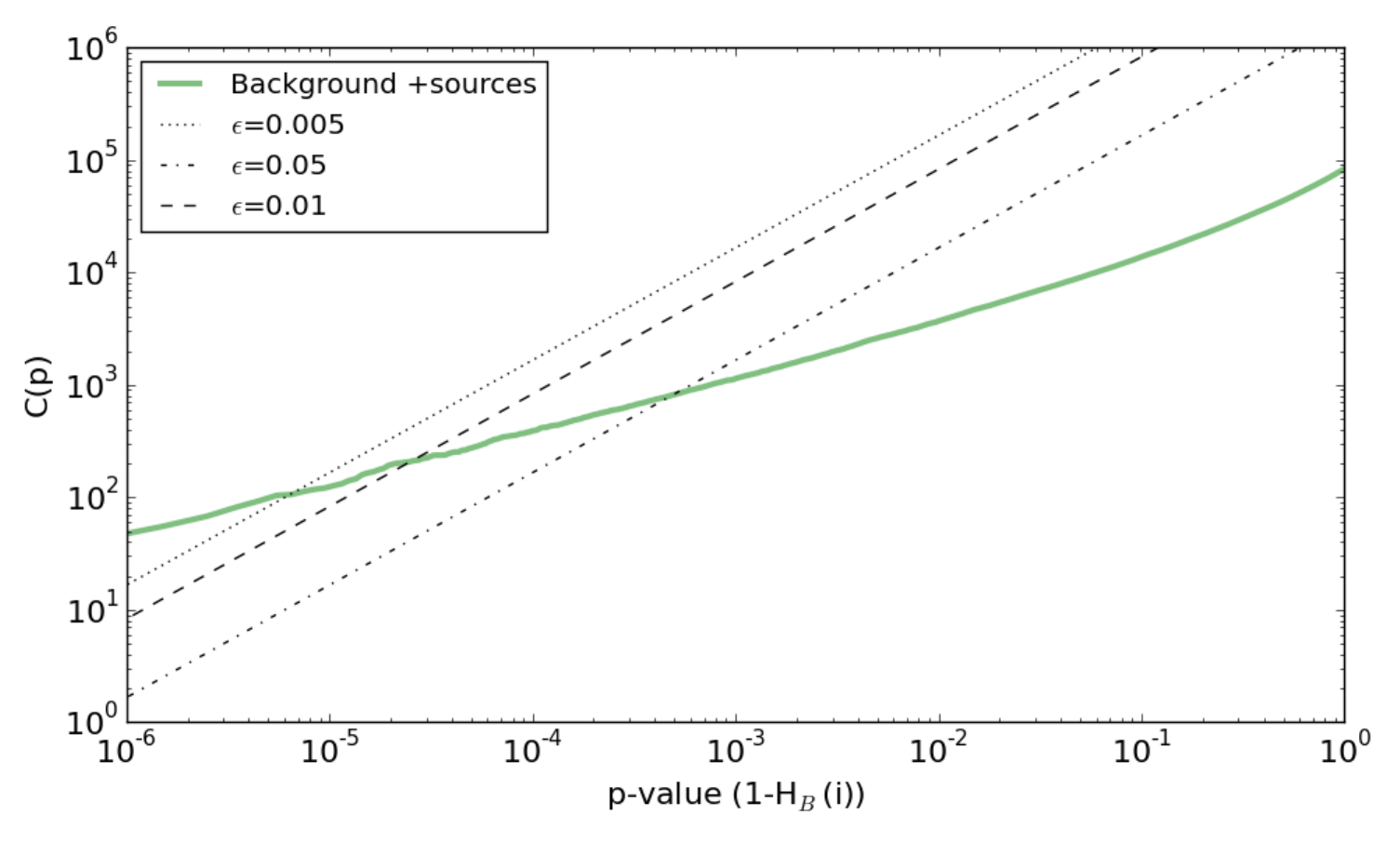}
\caption{\label{fig:cluster_selection}{\bf Top panel} distribution of excess number of objects (within 1\farcm0) with respect to the local background for all galaxy candidates at $z>1.3$. The green dashed line shows the  best fit Gaussian distribution fitted for values of $\Delta N_{CC} < 7$, and the probability of observing a given excess number of objects around a galaxy, under the null hypothesis that the galaxy does not belong to a cluster. {\bf Bottom panel} Illustration of the Benjamini--Hochberg procedure: the green solid line shows the cumulative distribution of galaxies' p-values computed under the assumption of the null hypothesis (i.e., using the Gaussian best--fit parameters). The black lines correspond to different levels of contamination of the final sample.}
\end{center}
\end{figure}

The observed distribution of excess number of objects (within 1\farcm0) with respect to the local background ($\Delta N_{CC}$) is shown in Figure~\ref{fig:cluster_selection} (top panel), together with the best-fit Gaussian distribution computed using values of $\Delta N_{CC} <3$. The best fit  Gaussian distribution has a mean of  $0.45$ and a standard deviation of $\sigma=4.6$, consistent with the best--fit values obtained by \cite{2014ApJ...797..109R} on similar depth data, on more than ten times the area. The Gaussian function can be used to describe the probability of observing a given excess number of objects around a galaxy, under the null hypothesis (H) that the galaxy does not belong to a cluster. In order to identify only those galaxies located within clusters we proceed following the Benjamini--Hochberg procedure \citep[BH; ][]{bh1995}, which minimizes the false discovery rate at a level $\epsilon$. Briefly, for each galaxy we compute its p--value under the null hypothesis H. The p-values are ordered in increasing order and denoted by $p_1,\dots\, , p_{N}$. The cumulative distribution of the p-values is shown in Figure~\ref{fig:cluster_selection}, bottom panel. Notice that the value of C(p) corresponds to the index $j$ of each galaxy (the galaxies were sorted according to their p-value). For a given $\epsilon$,  we compute the critical p-value by finding the largest $j$ such that $p_{j} \leq \frac{k}{N} \epsilon$. The  corresponding $\Delta N_{CC}$ is then the cutoff value we use to identify galaxies belonging to a cluster. The BH procedure ensures that the false discovery rate is smaller than $(\epsilon \times 100)$\%. In Figure~\ref{fig:cluster_selection} we show the curves corresponding to various values of $\epsilon$. For the cluster selection we used the conservative value $\epsilon = 0.005$, which corresponds to objects with $\Delta N_{CC} \geq 19.7$ (indicated by the vertical dotted line in the top panel of Figure~\ref{fig:cluster_selection}). 

\begin{figure}[ht!]
\begin{center}
\includegraphics[width=7.5cm]{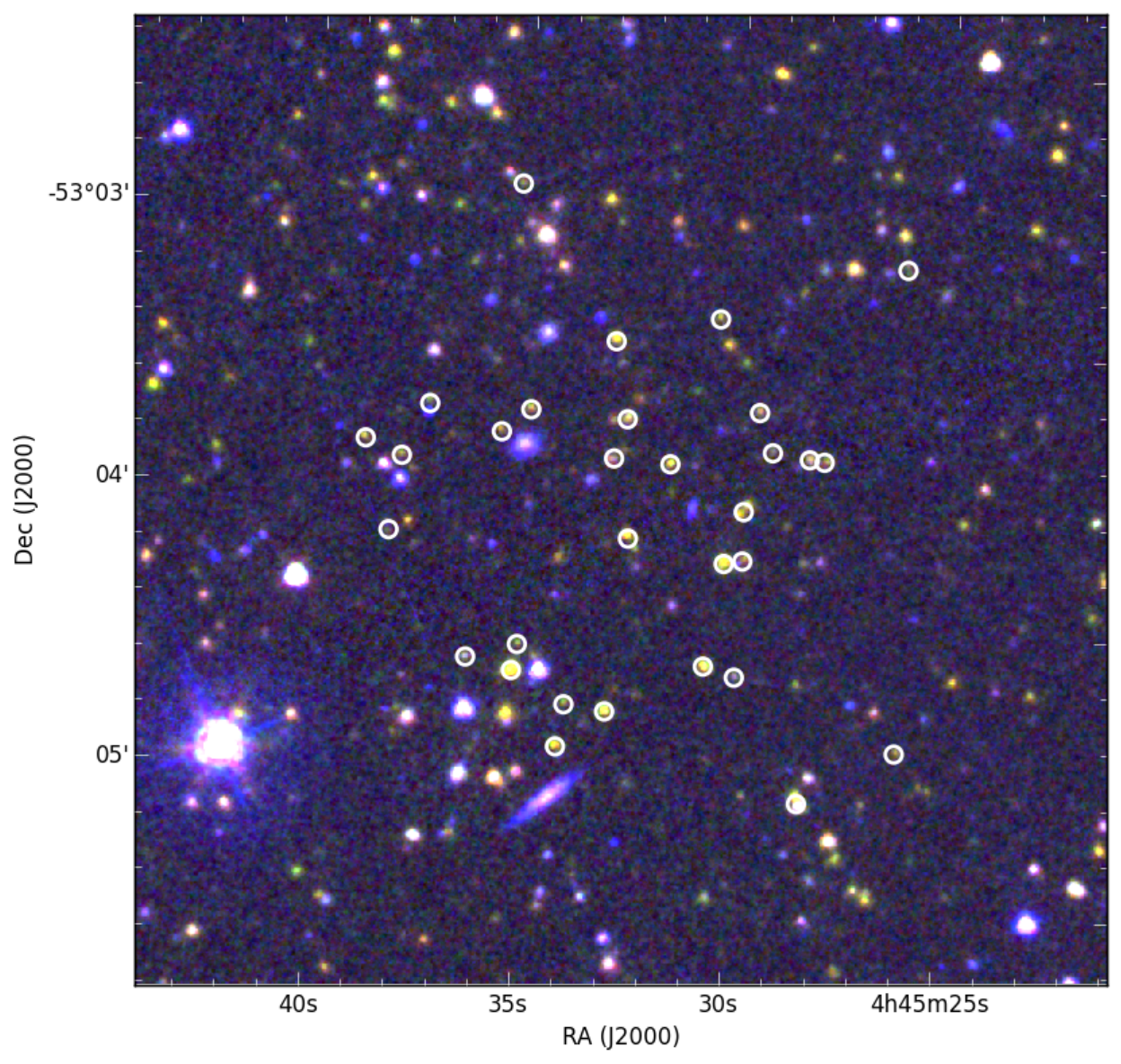}
\caption{\label{fig:cluster_example} False color image (red$=4.5$ \mic, green$=3.6$ \mic, and blue$=R_{c}$) showing a 3\farcm7$\times$3\farcm7 region around one of the galaxy cluster identified in the central square degree where optical data are available.  Open circles show galaxies with red 3.6$-$4.5 \mic\ color, 3.6 \mic\ and R$_{c}$ magnitudes fainter than 18.7 and 21, respectively.}
\end{center}
\end{figure}

The procedure above identifies galaxies residing in over dense regions, and thus, can identify multiple galaxies belonging to the same over density. We follow \cite{2008ApJ...676..206P} and \cite{2014ApJ...797..109R} and merge the cluster candidates by applying a friends-of-friends algorithm with a linking length of  1\farcm5, corresponding to approximately 0.8 Mpc at $z=1.5$. With this algorithm, we identify 27 unique galaxy clusters. An example of a detected cluster in the central region where optical data are available is shown in Figure~\ref{fig:cluster_example}. The density of clusters in the SIMES area (3.8$\pm$0.7 clusters deg$^{-2}$) is consistent with the density found by Rettura et al. (2014, 3.0$\pm$0.2  clusters deg$^{-2}$).

\section{SUMMARY}
We presented the \emph{Spitzer}-IRAC/MIPS Extragalactic survey (SIMES) in the South Ecliptic Pole field (SEP) and the multi-wavelength catalog of sources based on the 3.6 \mic\ detections. The survey covers an area of 7.74 deg$^{2}$ to a depth of $\sim$5.80 $\mu$Jy (3$\sigma$) at 3.6 $\mu$m and 5.25 $\mu$Jy at 4.5 $\mu$m. We estimate 90\% and 50\% completeness levels in the 3.6 \mic\ band at 14 and 9 $\mu$Jy, respectively.

The SIMES region has been  targeted by numerous multiwavelengths surveys spanning  the UV  to the far IR and radio regimes. The addition of the \emph{Spitzer}-IRAC observations is crucial for computing reliable photometric redshifts and stellar masses for all galaxies detected in this region by the {\it Herschel} satellite. The IRAC observation presented here allowed us to identify the optical/IR counterparts of the starforming galaxies and AGNs detected at the far--IR wavelengths. We included in our multi-wavelength catalog the WFI--R$_{c}$, MIPS--24 $\mu$m, SPIRE 250, 350, and 500 $\mu$m fluxes of the counterparts that we identified by searching for the closest neighbor. The reliability (i.e. fraction of spurious detections introduced) of these associations is quantified through the indicator ``P'' that we computed for each MIPS and SPIRE detected source.  The possibility of a direct IRAC--SPIRE association is also discussed. 
The full catalog is available through the NASA/IPAC Infrared Science Archive.

We reported  3.6 $\mu$m galaxy and \emph{total} (galaxy and stars) number counts in the SIMES field and compared  them with literature results obtained in different fields. Below $F_{3.6\mu m}=10^{-4.0}$Jy our galaxy counts are more in agreement with \cite{2006A&A...453..397F} than with \cite{2004ApJS..154...39F}. Above $F_{3.6\mu m}=10^{-4.0}$Jy galaxy counts in the SIMES field are in between those of \cite{2004ApJS..154...39F} and \cite{2006A&A...453..397F}.  
While our \emph{galaxy} number counts are computed within the area with optical imaging, our \emph{total} counts are calculated in the whole SIMES area.

Finally, using the method proposed in \cite{2010ApJ...716.1503P}, we identified 27 galaxy clusters at z$>$1.3. Although preliminary (only part of the field at this point is covered by optical data), the surface density of the galaxy clusters in SIMES is consistent with that reported in Rettura et al. (2014).   

Further deep observations in optical bands will be soon available (I. Baronchelli et al., in preparation). They will allow improvement of our estimates and the measure of precise photometric redshifts for the galaxies in the SIMES field. The correlation among the near-- and far--IR bands will be further improved using the available 90 $\mu$m Akari data presented in \cite{2014A&A...562A..15M}. These data will also allow for an extensive study of the dust thermal emission in these spectral regions.

\section*{ACKNOWLEDGEMENTS}
We thank the referee for the useful comments that improved the presentation of the paper.
I.B. and G.R. acknowledge support from ASI (Herschel Science Contract 2011aI/005/011/0).
C.S. and I.B. acknowledge support from NASA JPL/Spitzer grant RSA 1449911 provided for the SIMES project.
S.M. acknowledges financial support from the Institut Universitaire de France (IUF), of which she is senior member. 
M.V acknowledges support from the European Commission Research Executive Agency
(FP7-SPACE-2013-1 GA 607254) and the Italian Ministry for Foreign Affairs and
International Cooperation (PGR GA ZA14GR02).
This research has made use of data from HerMES project (http://hermes.sussex.ac.uk/). HerMES is a Herschel Key Program using Guaranteed Time from the SPIRE instrument team, ESAC scientists and a mission scientist. The HerMES data were accessed through the Herschel Database in Marseille (HeDaM - http://hedam.lam.fr) operated by CeSAM and hosted by the Laboratoire d'Astrophysique de Marseille. This research has made use of the NASA/IPAC Infrared Science Archive, which is operated by the Jet Propulsion Laboratory, California Institute of Technology, under contract with the National Aeronautics and Space Administration.
M.V. acknowledges support from the European Commission Research Executive Agency
(FP7-SPACE-2013-1 GA 607254) and the Italian Ministry for Foreign Affairs and
International Cooperation (PGR GA ZA14GR02).


\bibliographystyle{apj}
\bibliography{biblio1_r1.bib}{}

\begin{deluxetable}{lll}
\tabletypesize{\footnotesize}
\tablecolumns{3}
\tablewidth{0pc}
\tablecaption{Multiwavelength Catalog Columns and Description\tablenotemark{a}}
\tablehead{
\colhead{Column}& \colhead{Example} & \colhead{Content} }
\startdata
ID & 163210 & Identification number for IRAC 1 detected sources. \\
RA\_I1 & 71.040682 & IRAC 3.6 \mic\ RA coordinate \\
DEC\_I1 & -53.615640 & IRAC 3.6 \mic\ DEC coordinate \\
FLUX\_I1 & 4.58311 & IRAC 3.6 \mic\ total mJy flux (FLUX\_AUTO) \\
FLUXERR\_I1 & 0.00820884 &  IRAC 3.6 \mic\ mJy flux associated uncertainty \\
N\_SIGMA & 813.635 & IRAC 3.6 \mic\ Signal to noise ratio connected to average coverage\\
FLUX\_I2 & 3.05153 & IRAC 4.5 \mic\ total mJy flux (FLUX\_AUTO, I1 prior position used) \\
FLUXERR\_I2 & 0.00781307 & IRAC 4.5 \mic\ mJy flux associated uncertainty\\
RA\_24 & 71.040554 & RA coordinate for MIPS 24  \mic\ sources corrected for\\ 
& & systematic shift (-0\farcs165) \\
DEC\_24 & -53.615536 & DEC coordinate for MIPS 24  \mic\ sources corrected for\\ 
& & systematic shift (-0\farcs489) \\
O\_RA\_24 & 71.040600 & original RA coordinate for MIPS 24  sources \tablenotemark{b}  \\
O\_DEC\_24 & -53.615400 & original DEC coordinate for MIPS 24 sources \tablenotemark{b} \\
FLUX\_24 &  3.98400 & MIPS 24 \mic\ mJy flux \tablenotemark{b} \\
FLUXERR\_24 & 0.0600000 & MIPS 24 \mic\ mJy flux uncertainty \tablenotemark{b}\\
FLUX\_70 & 20.0000  & MIPS 70 \mic\ mJy flux \tablenotemark{b}\\
FLUXERR\_70 & 3.63636  & MIPS 70 \mic\ mJy flux uncertainty \tablenotemark{b}\\
RA\_SPIRE & 71.039836 & RA coordinate for SPIRE sources corrected for\\ 
& & systematic shift (-0\farcs199) \\
DEC\_SPIRE & -53.615458& DEC coordinate for SPIRE sources corrected for\\ 
& & systematic shift (-0\farcs443) \\
O\_RA\_SPIRE & 71.039856 &  original RA SPIRE coordinate \tablenotemark{c}\\
O\_DEC\_SPIRE & -53.615330 & original DEC SPIRE coordinate \tablenotemark{c}\\
FLUX\_250 & 92.635274 & SPIRE 250  \mic\ mJy flux \tablenotemark{c}\\
FLUXERR\_250 & 2.3467732 & SPIRE 250 mJy flux uncertainty \tablenotemark{c}\\
FLUX\_350 & 38.933088 &  SPIRE 350  \mic\ mJy flux \tablenotemark{c}\\
FLUXERR\_350 & 4.4716398 & SPIRE 350 mJy flux uncertainty \tablenotemark{c}\\
FLUX\_500 & 20.702205 & SPIRE 500  \mic\ flux \tablenotemark{c}\\
FLUXERR\_500 & 4.0425355 & SPIRE 500 mJy flux uncertainty \tablenotemark{c}\\
RA\_OPT & -53.615583 & RA coordinate for WFI--R$_{c}$\\
DEC\_OPT & -53.615583 & DEC coordinate for WFI--R$_{c}$\\
FLUX\_R\_WFI & 3.87101 &  WFI--R$_{c}$ total mJy flux (FLUX\_AUTO)\\
FLUXERR\_R\_WFI & 0.00968389 & WFI--R$_{c}$  total mJy flux uncertainty\\

P1 & 0.84638566 & SPIRE-MIPS reliability indicator \\
& & (ranges from 0 to 1, where 0=bad, 1=good)\\
N\_MIPS\_SPIRE & 1 & Number of MIPS counterparts for the SPIRE source\\
P2 & 0.85814963 & MIPS-IRAC reliability indicator \\
& & (ranges from 0 to 1, where 0=bad, 1=good)\\
N\_IRAC\_MIPS & 1 & Number of IRAC counterparts for the MIPS source\\
P3 & 0.80220842 & SPIRE-IRAC reliability indicator \\
& & (ranges from 0 to 1, where 0=bad, 1=good)\\
N\_IRAC\_SPIRE & 1 & Number of IRAC counterparts for the SPIRE source\\
CLASS\_STAR\_I1 & 0.0286267 & \emph{SExtractor} CLASS\_STAR parameter for IRAC 3.6 \mic\ \\
&&(ranges from 0 to 1, where 0=galaxy, 1=star)\\
A\_I1 & 6.28684 & Semi--major axis in arcseconds \\
B\_I1 & 3.39138 & Semi--minor axis in arcseconds \\
SIGMA &  0.00145476 & IRAC 3.6 \mic\ sky sigma value (depends on the coverage)\\
COVERAGE & 9.66118 & Average coverage computed over an area of 49 pixels \\
& & centered on the 3.6 \mic\ coordinates\\
AP1\_FLUX\_I1 & 0.00118364 & IRAC 3.6 \mic\ mJy aperture flux (4\farcs8 ap. diameter).\\ 
AP1\_FLUXERR\_I1 & 0.00118364 & IRAC 3.6 \mic\ mJy aperture flux uncertainty (4\farcs8 ap. diameter).\\ 
AP2\_FLUX\_I1 & 0.00184539 & IRAC 3.6 \mic\ mJy aperture flux (7\farcs2 ap. diameter).\\ 
AP2\_FLUXERR\_I1 & 0.00184539 & IRAC 3.6 \mic\ mJy aperture flux uncertainty (7\farcs2 ap. diameter).\\ 
AP3\_FLUX\_I1 & 0.00291065 & IRAC 3.6 \mic\ mJy aperture flux (12\farcs0 ap. diameter).\\ 
AP3\_FLUXERR\_I1 & 0.00291065 & IRAC 3.6 \mic\ mJy aperture flux uncertainty (12\farcs0 ap. diameter).\\ 
AP1\_FLUX\_I2 & 0.00118364 & IRAC 4.5 \mic\ mJy aperture flux (4\farcs8 ap. diameter).\\ 
AP1\_FLUXERR\_I2 & 0.00118364 & IRAC 4.5 \mic\ mJy aperture flux uncertainty (4\farcs8 ap. diameter).\\ 
AP2\_FLUX\_I2 & 0.00184539 & IRAC 4.5 \mic\ mJy aperture flux (7\farcs2 ap. diameter).\\ 
AP2\_FLUXERR\_I2 & 0.00184539 & IRAC 4.5 \mic\ mJy aperture flux uncertainty (7\farcs2 ap. diameter).\\ 
AP3\_FLUX\_I2 & 0.00291065 & IRAC 4.5 \mic\ mJy aperture flux (12\farcs0 ap. diameter).\\ 
AP3\_FLUXERR\_I2 & 0.00291065 & IRAC 4.5 \mic\ mJy aperture flux uncertainty (12\farcs0 ap. diameter).\\ 
\enddata

\tablenotetext{a}{All the fluxes are expressed in mJy and the coordinates in degrees. All the fluxes  are ``\emph{total''} and do not need any further aperture correction, unless differently specified. The IRAC aperture fluxes reported here for the 4\farcs8, 7\farcs2, and 12\farcs0 diameter apertures, are not aperture corrected; to obtain the correspondent \emph{total} flux, the IRAC handbook aperture corrections are needed. The counterpart distances are expressed in arc seconds. The catalog is released through the NASA/IPAC Infrared Science Archive (IRSA) service. }
\tablenotetext{b}{From \cite{2011MNRAS.411..373C}}
\tablenotetext{c}{From SPIRE XID catalogs \citep[DR2, ][]{2010MNRAS.409...48R,2012MNRAS.419..377S,2014MNRAS.444.2870W}}
\label{catalog_cont_table}
\end{deluxetable}

\end{document}